\documentclass{emulateapj}

\def\kms{~km~s$^{-1}$\ }
\def\kmsc{~km~s$^{-1}$}
\def\simgreater{\buildrel > \over \sim}
\def\simless{\buildrel < \over \sim}
\def\arcs{\char'175\ }
\def\arcsec{\char'175 }

\def\hub{\ifmmode H_\circ\else H$_\circ$\fi}

\shorttitle{Star Clusters in M31. II}
\shortauthors{Caldwell et al.}

\usepackage{color}
\usepackage{afterpage}

\begin{document}

\title{Star Clusters in M31: II. Old Cluster Metallicities and Ages from
Hectospec Data}

\author{Nelson Caldwell} 
\affil{Center for Astrophysics, 60 Garden Street, Cambridge, MA 02138, USA
\\ electronic mail: caldwell@cfa.harvard.edu}

\author{Ricardo Schiavon}
\affil{Gemini Observatory, 670 North A'ohoku Place , Hilo, HI 96720, USA \\ electronic mail:
rschiavo@gemini.edu}

\author{Heather Morrison}
\affil{Department of Astronomy,
Case Western Reserve University, Cleveland OH 44106-7215, USA
\\ electronic mail: heather@vegemite.case.edu}

\author{James A. Rose}
\affil{Department of Physics and Astronomy, University of North Carolina,
  Chapel Hill, NC 27599, USA \\ electronic mail:
 jim@physics.unc.edu}

\author{Paul Harding}
\affil{Department of Astronomy,
Case Western Reserve University, Cleveland OH 44106-7215
\\ electronic mail: paul.harding@case.edu}


\begin{abstract}
We present new high signal$-$to$-$noise spectroscopic data on the M31 globular cluster (GC) system, obtained
with the Hectospec multifiber spectrograph on the 6.5m MMT.  More than 300 clusters
have been observed at a resolution of 5\AA \ and with a median S/N of
75 per \AA , providing velocities with a median uncertainty of 6 \kms. The primary focus
of this paper is the determination of mean cluster metallicities, ages and reddenings.
Metallicities were estimated using a calibration of Lick indices with [Fe/H] provided
by Galactic GCs.  These match well the metallicities of 24 M31 clusters determined
from HST color$-$magnitude diagrams, the differences having an rms of 0.2 dex.  The metallicity
distribution is not generally bimodal, in strong distinction
with the bimodal Galactic globular distribution. 
Rather, the M31 distribution shows a broad peak, centered at  [Fe/H]=$-1$, possibly with minor peaks
at [Fe/H]=$-1.4$, $-0.7$ and $-0.2$, suggesting that the cluster systems 
of M31 and the Milky Way had
different formation histories.
Ages for clusters with [Fe/H] $> -1$ were determined
using the automatic stellar population analysis program {\it EZ\_Ages}. 
We find no evidence for massive clusters in M31 with intermediate ages, those between 2 and 6 Gyr. Moreover, we find
that the mean ages of the old GCs are 
remarkably constant over about a decade in metallicity ($-0.95
\simless $ [Fe/H]$ \simless 0.0$).

\end{abstract}


\section{Introduction}
Star clusters in the Andromeda galaxy
have been studied since \citet[][]{hubble}, and have long been known to
comprise ages from  young to old.  It is the latter
which are the topic of this paper.  Initially, cataloging
was all that was possible for clusters, work which indeed 
continues to this day with searches for
distant clusters \citep{huxor}.
But spectroscopic studies as a means of measuring  stellar populations
and group kinematics also began early, starting with \cite{vdb},
with substantial contributions by \cite{huchra82}, \cite{burstein}, \cite{trip},
\cite{huchra}, and \cite{federici}.
\cite{barmby} added much spectroscopy and photometry to the sample, and presented the largest study
before the use of multiobject spectrographs on M31.    
At the time of the Barmby et al (2000) work, it was 
generally thought that the old M31 globular clusters (GCs) numbered 
several hundred, spanned a range of metallicity
(as deduced from colors and absorption line
strengths) equal to
that of the galactic GCs, and likewise did not have a simple single Gaussian distribution
of metallicities. The metal$-$poor clusters were thought to be roughly spherically distributed
and to show a small amount of systemic rotation, while the most metal$-$rich clusters
were  thought to be 
confined to the disk of M31, and to show more systemic rotation, although the specifics
of those and subsequent results will be further refined in this paper and other papers
in this series.

Two large multifiber studies of M31 clusters have been presented in the last decade:
\cite{perrett} who used WYFFOS on the WHT, and \cite{kim} who used Hydra on the 
WIYN telescope. 
The former paper produced more than 200 new velocities, and 
also found a bimodality of
the old cluster metallicities,  using spectral indices rather than colors.
A subsequent kinematic analysis of those data by \cite{morrison} suggested that
the globulars could be explained using two kinematic components: a thin, cold rotating
disk and a higher velocity dispersion component whose properties resemble M31's bulge.
The second study \citep[where the kinematic results were presented in][]{lee} 
provided spectra for an additional 150 objects.
However, in both studies, the kinematic analysis suffered 
from the inclusion of
young disk clusters,  which in particular led to statements that
there were metal$-$poor clusters with thin disk kinematics \citep{morrison} or
that the metal$-$poor clusters showed strong systemic rotation \citep{lee}, in conflict with
what \cite{huchra} had reported, and what we will also conclude in a subsequent paper
(A. Romanowsky et al., in preparation).

The important task of distinguishing M31's young clusters from old was one of the
topics in
\citet[Paper I]{PaperI}, made possible with new spectroscopy 
taken on  the 6.5m MMT and using the Hectospec multi$-$fiber spectrograph \citep{fab}.
Ages and masses for more than 140 young clusters were determined by comparison with 
models, finding 
clusters with masses as great as $10^4 M_\sun$,  and a
median cluster age
of 0.25 Gyr.  Table 1 of that paper also listed all the
clusters, regardless of age, for which we added new
information to the long$-$standing cataloging efforts by \cite{galleti2}.  That new 
information
included revised coordinates, magnitudes, reddenings, 
and cluster classifications based on images and spectroscopy.

With the distinction between the young and old clusters now better clarified, 
we present here the first results of our high signal$-$to$-$noise (S/N)
spectroscopic study of the old clusters in M31 (where we define
old to be those with ages greater than 6 Gyr).  
This paper presents the velocities, ages and
metallicities for the old clusters, and discusses the
spatial, abundance and age properties of these clusters.
The improved data allow us to revisit 
the topics
addressed by previous authors.
Where possible, we present comparisons
with metal abundances derived from cluster color$-$magnitude diagrams (CMDs) using {\it HST} imaging, 
some of which is presented here.
Subsequent papers will discuss kinematics and
abundance ratios of these clusters. 

We assume a distance of 770 kpc throughout \citep{freedman}.

\section{Spectroscopy}

\subsection{Sample}

Subsequent to the publication of Paper I, more {\it HST} images of M31 clusters became available, which
were inspected closely 
to further aid in the classification of objects in the cluster catalog. These images 
were particularly useful for objects in the bulge of M31, because the ground$-$based images \citep{massey}
were often saturated in that high surface brightness area. 
There were no objects  previously
called clusters that are now thought to be stars, but several objects classified as stars based
on spectra alone have been verified to be stars from these images. 
The Web site {\it http://www.cfa.harvard.edu/oir/eg/m31clusters/\\M31\_Hectospec.html}
contains images for all old clusters, young clusters, stars and background galaxies
in the catalog of Paper I, and may be profitably used to view individual cluster
images as an adjunct to the spectroscopy of this paper. Spectra of the young clusters may
also be found there; those of the old clusters will also appear there upon publication 
of this paper.

The observations we report on here are comprised of objects called ``old'' in Table 1 of Paper I;
indeed those classifications were based on the information to be presented here.  

\subsection{Hectospec Observations}

The data were all taken at the MMT with the Hectospec multifiber spectrograph
\citep{fab} during the years 2004$-$2007. Twenty$-$five different but overlapping
1\degr \ diameter fields were observed, providing coverage except for
an area due west of the center  \citep[Figure 2]{PaperI}.  Each field had targets of various
kinds: clusters, HII regions and planetary nebulae, but the exact kind of cluster being observed, 
whether old or young,
was often uncertain until the data were analyzed.
Exposure times were
typically 3600$-$4800s. Of the 
367 clusters contained in our catalog \citep{PaperI} that we thought were old,
we report on new spectra of 346, all but 26 of which we maintain are
indeed old.  We also  report briefly on more than 30 clusters that
appear to be younger than 2 Gyr upon further
examination of their spectra, and which should have appeared in Table 2 of Paper I. 
A few cataloged old clusters within 1.\arcdeg6
of M31's center were missed 
in the spectroscopic observations either because they were
outside of the area surveyed (the massive cluster G001 was missed for this reason), 
or initially had bad coordinates and were not re-observed
after their coordinates had been corrected.  There are as well several 
candidate clusters
from \cite{kim} that may be old clusters, judging from their ground-based images, but for which we
have no spectra. These are concentrated in the disk of M31. Finally, aside of one case,
we did not observe the
outer clusters that have recently been discovered and cataloged by \cite{huxor} and
succeeding papers.  The Hectospec fields covered a radius of about 1.6\degr = 21 kpc.
Within that  circle, there are 339 old clusters;  we observed 316
of those, for a completeness of 93\%.

Our spectra cover the wavelength range of 3700$-$9200 \AA \ at 5\AA \ resolution, with
dispersion 1.2\AA \ pixel$^{-1}$, 
and were reduced as described in Paper I.  The fibers are 1.5\arcs in diameter. 
Thus an important distinction needs to be made concerning the fraction  of total
cluster light these spectra represent, and the amount contained in typical spectra
of galactic GCs. According to \cite{barmby07}, the half-light radii of M31 GCs
range from 0.5 to 1.0\arcs. Thus, the Hectospec data sample the cluster light very
well for most of the clusters. The Milky Way (MW) GC spectra by contrast sample mainly the
cores of those clusters \citep{schiavon05}.

Sky subtraction
was performed using several object-free spectra as near as possible to each target, except
for targets in the bulge area, where the local background was high. In those cases,
only sky spectra
far from the bulge of M31 were used.  A separate offset exposure for such  fields, 
taken concurrently and about 5\arcs offset from the targets, was reduced in 
a similar way (so that contemporaneous sky subtraction was performed for on- and
off-target exposures), and then the off-target
spectra were  subtracted from the on-target.  Tests done using spectra of the same bulge clusters
reduced with and without the local background subtraction revealed errors of up to 100\kms
and systematic errors of order 20\kms, if a local background was not used.
The errors were in the sense that the observed velocities were 
pulled toward the velocity of the local background.  Thus for this project 
local background measurements were crucial for obtaining accurate
results,  and this paper is the first to present precise
  velocities for a large number of clusters projected on the inner few kpc of M31.

\begin{figure}
\vspace{1.0cm}
\plotone{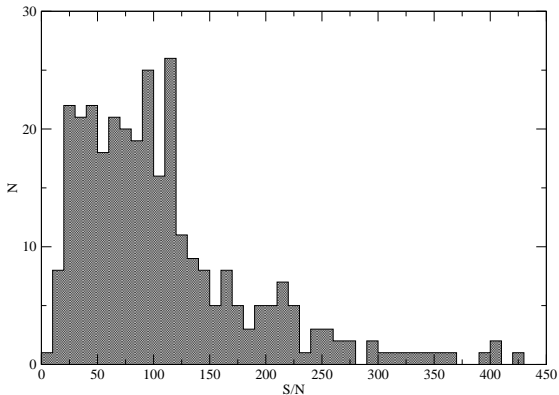}
\caption{\small Distribution of S/N at 5200 \AA \ for 316 GC spectra. \label{snr}}
\end{figure}

The
median S/N at 5200\AA \ of the main set of spectra 
is 75 per \AA, with some as high as 300 and others as low as 8 (Figure \ref{snr}).  
The spectra were corrected to relative fluxes using 
standard star measurements taken at irregular intervals during the
observation period.  The spectrograph throughput was remarkably consistent
over this period \citep{fab2}, thus we are confident in the corrections to fluxed
spectra. The fluxing process was needed only for determining the
reddening values. 

A small number of clusters were observed with a higher ruling grating, which 
gave spectra with a resolution of 2.1\AA \ over the range of $6000-8500$\AA.
These spectra are useful for velocities, but not for analyzing the stellar populations.
Therefore, the ages of these clusters are still unknown.

\subsection{Velocity Measurements}
Subsequent to the Hectospec project, a HectoChelle project 
of many of these
same targets was initiated \citep{strader}, resulting in more precise velocities which 
allowed us to refine the Hectospec
velocity zero$-$points. The HectoChelle spectra have a resolution of 30,000
and mean internal errors less than 1 \kms \citep{saint} . 
In order to set the velocity zero$-$point for the Hectospec
  data, we first  wavelength calibrated, and then 
set the velocities for the Hectospec spectra 
of  149 objects in common to be equal to
the values found from the HectoChelle spectra. (There are 186 old clusters
in total for which we have good HectoChelle spectra).  These Hectospec spectra were then
shifted and rebinned to zero$-$velocity, and combined into a single template.
That template was then used as the cross$-$correlation  template for the full
set of Hectospec spectra, using the SAO xcsao software \citep{kurtz}.  
By doing this, we minimized zero$-$point errors
due to stellar template velocity
errors and coarse spectral type mismatches that can occur when using a single
star template on a composite spectrum.  We did however include the entire
range of metal abundances
for the M31 GCs to make the template, thus there may still be some residual 
velocity errors due to template mismatching, for instance in a case where
the cluster spectrum is very different from the mean cluster spectrum. The
HectoChelle data (see below) proved to be a useful check on systematic errors.

The cross$-$correlation analysis for the Hectospec spectra used wavelengths from
4000-6800 \AA, excluding wavelengths from 3700-4000 \AA \ (the H\&K region)
even though
the spectra cover that bluer region. 
If we incorporated that spectral region, we found a moderate dependence of velocity
difference with velocity; as compared with the HectoChelle velocities. The trend
was 20 \kms over the range of 0 to $-$700 \kms. This was  probably
the result of wavelength calibration errors for the bluest wavelengths. 
With the bluest spectral region excluded, 
the trend was reduced to 5 \kms. The median offset in 
velocity became +2.9 \kms (Spec$-$Chelle) and the rms became 5.7\kms.
We did not remove this small residual offset from the velocities reported here.

Table \ref{main} lists the objects believed to be old globular clusters
from our sample, along with their coordinates, velocities and uncertainties. The
other data presented in this table are discussed below.  The velocities in this table come solely from
the Hectospec data (with 3 exceptions where we had only HectoChelle data); 
the HectoChelle velocities will be reported  in \cite{strader}.
Table \ref{revised} lists clusters
thought to be old in Paper I, but which are now realized to be younger than
2 Gyr. Their velocities and uncertainties are also reported.  The ages were determined as in Paper I,
and again have uncertainties of about a factor of 2.

\begin{figure}
\vspace{1.0cm}
\plotone{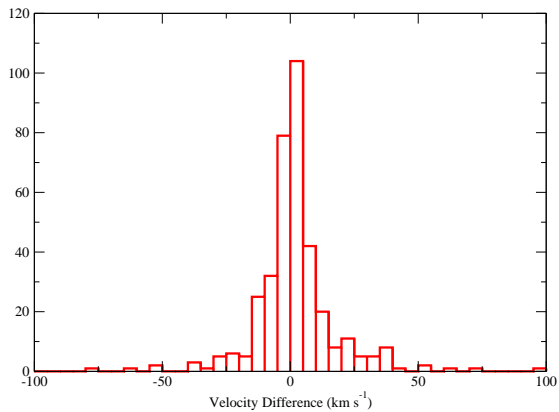}
\caption{\small Distribution of velocity differences derived from repeated measurements
of 213 M31 GCs. The median of the  velocity differences is 1.2 \kmsc, with an rms=8.8.
This implies a median single measurement uncertainty of 6.2 \kmsc. \label{repeat}}
\end{figure}

We can assess the quality of the  Hectospec velocities in three ways: internally via repeated
measurements, externally with the HectoChelle velocities, and externally with
literature values.
We have repeat observations for 213 GCs, some more than once for a total of 369
measurements. The median of the  velocity differences is 1.2 \kms, with an rms=8.8.
This implies a median single measurement uncertainty of 6.2 \kms. Figure \ref{repeat}
shows a histogram of these velocity repeat differences.

\begin{figure}
\vspace{1.0cm}
\plotone{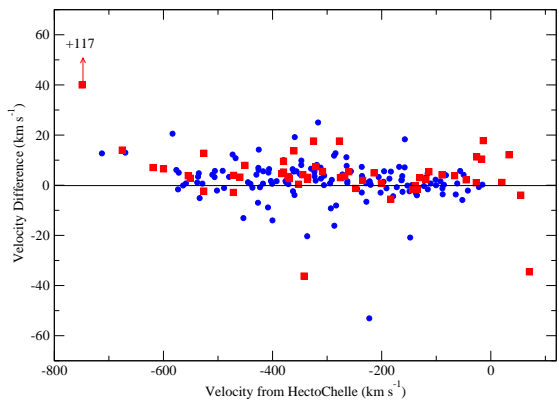}
\caption{\small Comparison with HectoChelle measurements. Velocity differences in the sense
Hectospec - HectoChelle are plotted against the HectoChelle velocity.
Squares indicate clusters in the bulge.
The median offset in 
velocity is +2.9 \kms (Spec$-$Chelle) and the rms is 5.7\kms.
The worst case is that of B070-G133,  which is just outside our defined bulge area, 
and thus its velocity may also be affected by bulge contamination. \label{chelle}}
\end{figure}

Figure \ref{chelle} shows the velocity differences of the 187 GCs that 
have HectoChelle velocities with cross correlation coefficients higher than
7, plotted against
their velocities to allow the detection of velocity dependent 
errors.  
 Four of these GCs have discrepant velocities in the Hectospec spectra: B070-G133 differs
from the HectoChelle velocity 
by $-53$\kmsc, B132 by $-35$\kmsc,  B262 by  $-36$\kmsc, and NB21 by $+117$\kmsc. 
These errors probably resulted  because all four are 
in the bulge.  The first did not have an offset background exposure at all, 
and NB21 is the cluster closest to M31's nucleus for which we 
have a spectrum, and thus it is the most affected by contamination from
the surrounding light.
The rms of the other 183 velocity differences is 5.7\kmsc , whereas the
median formal uncertainty in HectoChelle velocities is 0.5 \kmsc . Thus, the 5.7 \kms is probably 
largely due to the Hectospec data,
and therefore is close to the true median external
uncertainty of  objects in the Hectospec velocity catalog. The median
Hectospec formal uncertainty from xcsao (the internal uncertainty)
 of those same objects is 9.8\kms, so the 
formal uncertainties listed in Table \ref{main} are somewhat of an overestimate of the
external uncertainties.  For objects with
listed uncertainties less than 20 \kms, the external uncertainties can be estimated as
$external=0.9+0.4*internal$. Figure \ref{chelle} also shows the remaining
zero$-$point offset of 3\kms and the remnant velocity$-$correlated velocity differences.
The HectoChelle velocities will be presented in  \cite{strader}.

There are two large collections of M31 GC velocities with which we can compare
our data, \cite{barmby} and \cite{perrett}.  
There are not enough velocities in common with those reported
by \cite{kim} to construct a meaningful comparison.
Figure \ref{compare_vels} shows a comparison of Hectospec velocities with those
two sources.  For the \cite{perrett} comparison, there were two cases where
the object called a cluster by that source was in fact a background galaxy; these
we have excluded from the comparison. In the Barmby comparison, we have
 186 objects in common, and the median velocity difference is $-$6 \kms, 
with an rms of 90 \kms. This is not too surprising given the heterogeneity of that
catalog. For the Perrett comparison, there are  185 in common, with a 
median difference of $-$1\kmsc, and an rms of 52\kmsc, dropping to 30 if the remaining four worst, 
all of which are bulge clusters, are thrown out. Perrett quoted an
overall uncertainty of 12\kmsc, thus we find their uncertainties to be underestimates.
Comparison of the Perrett velocities with the HectoChelle velocities allows us to
derive a median uncertainty of 26 \kms for their data set.

After putting them on a flux scale, the spectra were corrected to zero$-$velocity using our derived 
velocities, in preparation for index measurements  and reddening determination. 
\begin{figure}
\vspace{1.0cm}
\plotone{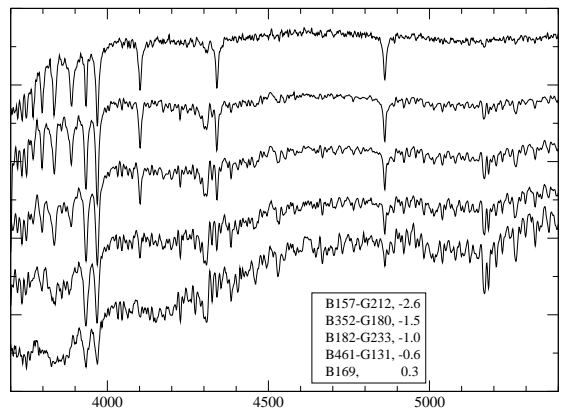}
\caption{Sample spectra from Hectospec.  The [Fe/H] values are listed
next to the objects' names, and 
are those derived in this paper.  The S/N ranges
from 20 per \AA \ (B169) to 100 (B182-G233)   at  4000\AA, and 50 to 200 at  
5200\AA. These spectra have been flux calibrated and Gaussian smoothed using a $\sigma =0.7$ \AA .
The full spectral coverage of 3650-9200\AA \ has been truncated for display purposes.
 \label{sp1}}
\end{figure}
Figure \ref{sp1} shows a sample of the spectra, covering
the full range of metallicities, which are described in the next section.

\section{Cluster Metallicities via Milky Way GC Calibration}

In this section  we will discuss the measurement of the
overall cluster metal abundances, characterized by [Fe/H], using
absorption line indices from the M31 GC spectra.  The indices can be converted to 
[Fe/H] using a calibration of the same indices found in the integrated 
spectra of  Milky Way (MW) GCs,  whose
metallicities have previously been 
measured through abundance analysis of high resolution spectra of
bona fide cluster member stars.
Using such a calibration implicitly assumes that all the clusters, M31 and MW,
are the same age. Thus we ignore not only the possibility that clusters in
the two galaxies span
a different range of ages, but also the observation that stars within
many MW clusters may themselves have a range of ages \citep[e.g.,][]{piotto}. 
The latter is very likely to be of negligible importance for the purposes of this
study, as the age differences are found to be in most cases exceedingly small,
or involve a small fraction of the total cluster's stellar population.
A larger difficulty is that the coarse Lick line indices
we employ have complicated behaviors with [Fe/H]. For instance, the Fe, Mg~b and
Balmer indices all have a break in their relations  at [Fe/H] values near
$-1.5$. Also, the Balmer indices become insensitive at the high metallicity
end,  [Fe/H] $> -1$, which of course makes them useful when combined
with an Fe index to determine ages and metallicities simultaneously, but
useless to determine metallicities for metal$-$rich clusters.
First, we will examine
the distribution of the indices themselves.

\subsection{Line Indices and Their Distribution }  

To measure the line indices, we used the 
lick\_ew code,  which is part of the 
EZ\_Ages\footnote{\small 
http://www.ucolick.org/$\sim$graves/EZ\_Ages.html} package \citep{ezages}.
We  smoothed the M31 spectra to the lower Lick/IDS
resolution (see \cite{worthey2} for details) and measured the
equivalent widths (EWs) adopting the passbands defined by \cite{worthey1}
and \cite{worthey2}.  The instrumental EWs were converted to the
Lick system \citep[as redefined by][]{schiavon07} using zero$-$points
determined from measurements taken in spectra of six Lick standards
also observed with Hectospec. These data,
taken typically through one fiber only,
were reduced following the same procedure as for the GC
spectra and provided small zero point shifts used to convert the instrumental
EWs into the Lick system. The details of the transformation to the standard
Lick system will be covered in a subsequent paper (Schiavon et al. 2010, in preparation).

\begin{figure}
\vspace{1.0cm}
\plotone{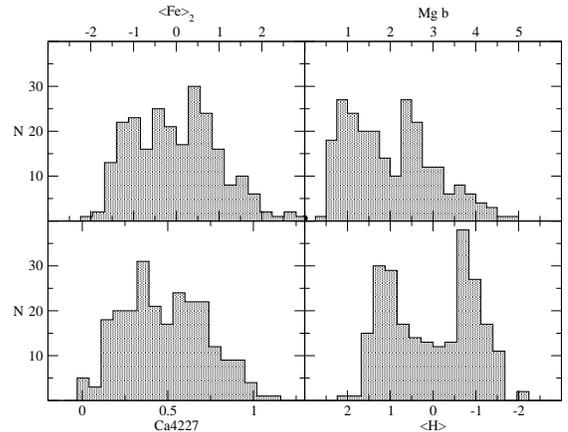}
\caption{Histogram of 4 different indices. The $\langle$Fe$\rangle_2$ and $\langle$H$\rangle$ indices are
described in the text.  Units are equivalent width in \AA \ for Ca4227 and Mg~b, standard deviations for
the other two.
Indices for 240 clusters with S/N $>$ 50 at 5200\AA \ are shown here.
Note that the X axis for $\langle$H$\rangle$ has
been reversed so that metallicity increases to the right as in the other index plots. \label{indices}}
\end{figure}
Figure \ref{indices} shows the distribution of several selected indices: 
$\langle$Fe$\rangle_2$, Mg~b, $\langle$H$\rangle$, and Ca4227.
The $\langle$Fe$\rangle_2$ index is the average of 2 Lick Fe indices :
 Fe5270 and Fe5335,
formed by converting the cluster values for each index to zero mean, unit variance values, and then
averaging. $\langle$H$\rangle$ is the average of H$\delta_{\rm F}$, H$\gamma_{\rm F}$  and H$\beta $ indices \citep[definitions
contained in][]{worthey1, worthey2}, formed in the same way.
The index histograms are roughly bimodal, but complex.
The conversions of these integrated light 
indices to [Fe/H] are not linear however, and thus the index
distributions are not simple indicators of the underlying
metallicity distribution, which we show below is neither unimodal
nor bimodal.

\subsection{ [Fe/H]  using Lick  Fe indices }
\label{s:emp_feh}
Early methods of ranking integrated light spectra of MW GCs by metallicities 
essentially used the similarity 
of the integrated spectra to
those of single stars, and typed the cluster spectra as one would type
a single star \citep{ mayall, morgan}. Strong Balmer lines due to the 
presence of A-F stars indicated
metal$-$poor clusters, while features due to G-K stars of course were found in metal$-$rich
ones.
Our method here for the M31 clusters 
derived from the methods of  \cite{dacosta} and \cite{brodie1990}, who measured
strong metallic lines in the spectra of MW GCs which had existing estimates for [Fe/H], and
used the correlations of the line indices with [Fe/H] to then derive [Fe/H] values for
extragalactic clusters.  The difference in our method is that with
our high S/N, we could make use of weaker lines that are more closely related to the
actual Fe abundance in the stars of the clusters. 

The disadvantage of our ignoring the stronger lines
was that the uncertainties for the metal$-$poor clusters were larger than
would have been the case had we considered such lines as the Balmer series
or the Mg~b lines.  \cite{galleti09} recently used
a heterogeneous set of spectra to estimate M31 GC metallicities, using
a combination of weak and strong spectral features.
Indeed, one can imagine other techniques that use more information
in the spectra, perhaps in a $\chi^2$ matching method in comparison with a set of
MW GC spectra, but given the incomplete set of  MW GC spectra, that method
is beyond the scope of this paper.  
\begin{figure}
\vspace{1.0cm}
\plotone{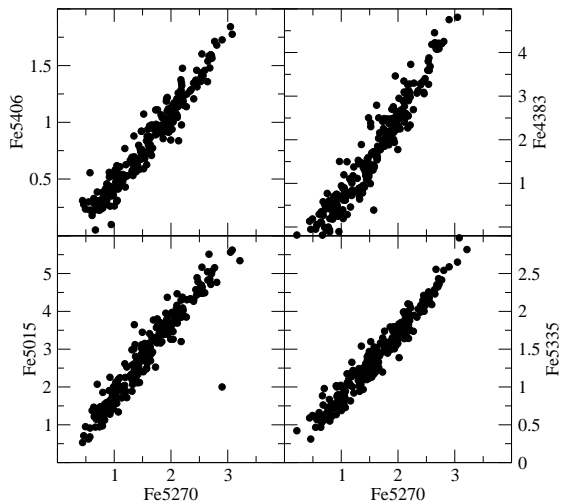}
\caption{Comparison of 5 Lick iron indices derived from the M31 GC spectra.  
Fe4383, Fe5015, Fe5335, and Fe5406 are plotted against Fe5270. The indices plotted
here are from spectra with S/N $>$ 30 at 5200\AA , typically the index 
uncertainties are less than 0.2\AA.  The discrepant point in the Fe5015 plot is B115-G177, which
has strong [OIII]$\lambda 5007$ emission affecting the index measurement.\label{feh indices}}
\end{figure}
Figure  \ref{feh indices} shows the mutual correlation of five Lick indices that
were originally designed to measure Fe in stellar populations, 
and indeed they all correlate very well with each other.  However,
we calibrated the Lick indices with [Fe/H] using spectra of MW GCs, and one
of those indices was not measurable using the available MW GC spectra. Also, including
Fe4383 and Fe5406 did not improve the errors in the derived [Fe/H].
Thus we restricted ourselves
to using an average, $\langle$Fe$\rangle$,  of just the  Fe5270 and Fe5335  indices to estimate [Fe/H].

The calibration was based on the library of MW GC spectra by
\cite[the ``Tololo'' sample,][]{schiavon05}.  This library consists of flux-calibrated,
intermediate-resolution, spectra of 41 Galactic GCs obtained with the
Blanco 4~m telescope at CTIO.  The spectral coverage is approximately
3350--6430 ${\rm\AA}$, and the spectral resolution $\sim$ 3.1
${\rm\AA}$.  Lick Fe indices were measured for the MW spectra in the same
manner as for the M31 spectra.

\begin{figure}
\vspace{2.0cm}
\plotone{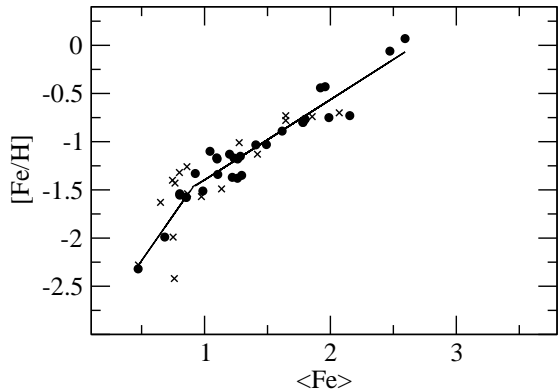}
\caption{Relation of the Lick $\langle$Fe$\rangle$ spectral index derived from the 
data of \cite{schiavon05} and [Fe/H]  for Galactic Globulars (circles). This
bi-linear
relation was used to derive the [Fe/H] values for the M31 clusters.  Crosses represent Lick
indices of additional Galactic Globulars  from
\cite{burstein}. These were not used in the fit, but serve to confirm the validity 
of the fit. \label{MW indices}}
\end{figure}

The iron abundances spanned by the MW GCs in the \cite{schiavon05} study cover
essentially the entire range of metallicities of the MW GC system ($-2.5<$
[Fe/H]$<0$) and are based on the metallicity 
scale of \cite{carretta}. 
Our relation between
$\langle$Fe$\rangle$ and [Fe/H] for 31 MW clusters with \cite{carretta} abundances
is shown in Figure \ref{MW indices}.
A bi$-$linear fit was deemed to be best, and avoided the problems of wild extrapolated
values that a quadratic or cubic fit would give.  The transformation we derived is: 

\noindent
 [Fe/H] = $ -2.23 + 0.83 * $ $\langle$Fe$\rangle$ ; for $\langle$Fe$\rangle$  $ > 0.9$; \\
\noindent
 [Fe/H] = $ -3.18 + 1.88 * $ $\langle$Fe$\rangle$ ; for $\langle$Fe$\rangle$  $ \le 0.9$.

Indices for different MW clusters reported in
\cite{burstein} are shown here as well to confirm the break in the relation (these were
not used in deriving the bi$-$linear relation).  Though we do not show them here, 
relations between [Fe/H] and Mg~b and Balmer indices also showed a break.
These breaks occurred at about [Fe/H]$=-1.2$ to $-1.5$, and like the one for $\langle$Fe$\rangle$, had
the effect of broadening the low metallicity side of the index distributions when they 
were converted into metallicity distributions.  Similar behavior was seen in
the relation between line indices and MW GC [Fe/H] values shown in \cite{galleti09}.
If instead we used a single linear relation for any of these relations, the result 
was an overestimate
of the metallicities of the most metal$-$poor clusters. Those with [Fe/H] $\simeq -2.5$ were
then estimated to have [Fe/H] $\simeq -2.0$ while those with [Fe/H] $\simeq -1.8$ were
then estimated to have [Fe/H] $\simeq -1.7$.

The MW GC bi$-$linear relation was then used to derive [Fe/H] values for the M31 clusters from
these Lick indices, which
are listed in  Table \ref{main}.  We allow an extrapolation in [Fe/H] to $+0.3$, beyond 
the last MW GC data  point at [Fe/H]$ \sim -0.1$.  For that reason, [Fe/H]
values beyond that limit are obviously less reliable, though we do use a linear
extrapolation.
(Table \ref{main} also  includes the few clusters thought to be old based on their images
that were observed only with the 600gpm grating centered at 
7000\AA . These were of course useful for velocities but not for the stellar population
analysis. Thus, they have no [Fe/H] values listed. These include B056D
and six clusters from the \cite{kim} catalog.)

\begin{figure}
\vspace{3.0cm}
\plotone{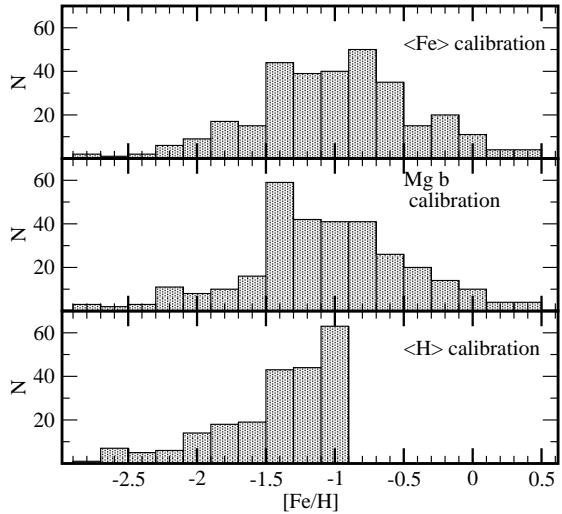}
\caption{Metallicity distribution of M31 GCs using three different calibration relations
for transforming an index to [Fe/H].  Values in the top plot were derived using
$\langle$Fe$\rangle$, those in the middle from Mg~b, and those in the bottom plot from
$\langle$H$\rangle$. Only the more metal-poor clusters are shown in the $\langle$H$\rangle$ panel, because
the calibration is ill-defined for the more metal-rich clusters.
This plot serves as an indication of the systematic
errors in the [Fe/H] values contained in Table \ref{main}. \label{hist_compare}}
\end{figure}

Now, the break in an index$-$[Fe/H] relation can transform the nature of the line index histogram
from clearly  bimodal to a more complex distribution, and this is what has 
apparently happened to our
M31 data using our MW GC
$\langle$Fe$\rangle$$-$[Fe/H] relation (see Figure \ref{hist_compare}, top panel), where the 
[Fe/H] histogram is not as obviously bimodal as is the $\langle$Fe$\rangle$ distribution.  
The GC color bimodality seen in most 
other external galaxies is taken to be
caused by an intrinsic bimodality in the GC metallicity distribution \cite[e.g.,][] {brodie}, but 
\cite{yoon} examined the relation of color and metallicity, and found that relation also to 
be non$-$linear. Thus Yoon et al. doubted the universality of bimodal metallicity
distributions in galaxies.

The uncertainties in the tabulated [Fe/H] values are derived from the statistical uncertainties 
in the $\langle$Fe$\rangle$ line indices
used and from the formal uncertainties  in the fit of [Fe/H] to the MW GC indices.  
Of course, there is an additional
systematic uncertainty due to the choice of [Fe/H] values for the  MW clusters 
used in the fit, the functional form of the fit that we chose, and the particular
indices we chose to fit to [Fe/H].
To characterize one aspect of the systematic uncertainties, 
we show a comparison of the metallicity distributions of the entire
M31 sample where the metallicities are derived from three different [Fe/H] calibrations, the
adopted  $\langle$Fe$\rangle$$-$[Fe/H] relation, and similar (bi$-$linear) relations involving Mg~b and a Balmer
index. 
These histograms are 
shown in Figure \ref{hist_compare}.  The basic difference is a stronger peak at [Fe/H]=$-1.4$ from
the Mg~b relation, but otherwise the agreement is good.  There is no systematic 
trend between the two
determinations, and excluding five outliers, the rms between them is 0.18 dex for 
316 clusters.
The data for the [Fe/H] derived from $\langle$H$\rangle$
are cut off at   [Fe/H]=$-1.0$ because the index has little sensitivity at higher metallicities.
Still, it is clear that the strong bimodality shown in the $\langle$H$\rangle$ index distribution
is not reflected in the derived
metallicity distribution.

We do not consider the calibration of the Lick indices with [Fe/H] to be
a solved question, though, and thus the resultant conclusions regarding 
the metallicity histogram (contained in  Section \ref{s:metallicity} below) are preliminary.
Clearly, more metal$-$poor MW GCs would
address some of the shortcomings of the calibrations we have employed here, but verification of [Fe/H]
values via individual stars for a representative sample of M31 GCs is also desirable.

\section{Metallicities and Ages via Population Synthesis}

Both ages and chemical composition estimates for the M31 GCs were obtained 
simultaneously from
comparisons of the Lick indices with the SPS models from \cite{schiavon07}. 
The details of the models are contained in that paper, but in brief,
polynomial functions describing the relations between 
various spectral indices and physical parameters of a stellar library
were computed, and then combined with theoretical
isochrones in order to produce predictions of integrated
indices of single stellar populations.
The isochrones employed
were those from the Padova group for both the solar$-$scaled 
\citep{girardi} and $\alpha$$-$enhanced cases \citep{salasnich}.

\begin{figure}
\vspace{1.0cm}
\plotone{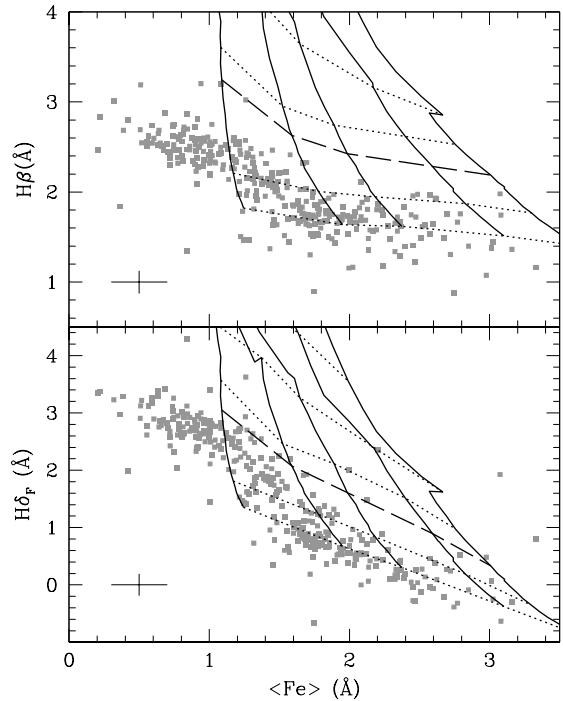}
\caption{  Comparison between data for M31 GCs and the \cite{schiavon07}
models.  Model same-[Fe/H] lines (solid), from left to right, are for [Fe/H] =
$-$1.3, $-$0.7, $-$0.4, 0.0, and +0.2, whereas same-age lines (dashed), from
bottom to top, are for 14, 8, 3.5, 2.5, 1.5, and 0.9 Gyr.  Both panels show
$\langle$Fe$\rangle$ (average of Lick indices Fe5270 and Fe5335) plotted against a Balmer
line: H$\beta$ (top) and H$\delta_F$ (bottom).  Average error bars are
shown in the bottom left corner of each panel.  Clusters with blue
horizontal branches (mostly with [Fe/H] $\lesssim -1.0$) have artificially 
young ages.  EZ\_Ages does not deal
with model extrapolations, and as a result, clusters outside the model grid are
excluded from the analysis (see text for discussion).}  \label{hbhdfem}
\end{figure}

Figure~\ref{hbhdfem} shows a comparison of data and models in two
index$-$index diagrams.  Both panels have the average $\langle$Fe$\rangle$ of the
Lick Fe5270 and Fe5335 in the abscissa, plotted against Balmer line
indices H$\beta$ and H$\delta_F$, in the top and bottom panels,
respectively.  Data points correspond to measurements of Lick indices
in the spectra of 316 M31 GCs.  Average error bars are displayed
in the lower left corner on both panels.  Because H$\beta$ and
$\langle$Fe$\rangle$ are relatively insensitive to abundance ratio variations
(for constant [Fe/H]), these indices provide reliable first guesses
on the ages and iron abundances of the clusters under analysis.  

Age and metallicity determination was achieved through application
of the EZ\_Ages code, developed by \cite{ezages} for automatic
stellar population analysis.  EZ\_Ages\footnote{\small See
http://www.ucolick.org/$\sim$graves/EZ\_Ages.html} is an IDL
implementation of a method developed by \cite{schiavon07} to estimate
the luminosity$-$weighted ages of stellar populations, as well as
their luminosity$-$weighted abundances of iron, magnesium, carbon,
nitrogen, and calcium, from Lick indices measured in their integrated
spectra.  The method has been successfully tested through application
to Galactic GCs with known ages and chemical composition.  Briefly,
the method relies on a sequential grid inversion algorithm, which
resolves for the parameters that fit best various index pairs,
starting with age and the abundance that affects the most observables
([Fe/H]), then going down hierarchically to estimate abundances
that impact fewer and fewer observables.  Thus, adopting a first
guess on the abundance pattern of the object stellar population,
EZ\_Ages finds the age and [Fe/H] combination that matches H$\beta$
and $\langle$Fe$\rangle$ best.  Because these two indices are relatively insensitive
to abundance$-$ratio variations, they provide a very good first guess
on age and [Fe/H].  Assuming the latter values, EZ\_Ages next finds
the [Mg/Fe] value that provides a best match to magnesium$-$sensitive
indices (Mg~$b$ and/or Mg$_2$).  Also adopting age and [Fe/H]
inferred from the match to the H$\beta$$-$$\langle$Fe$\rangle$ pair, EZ\_Ages finds
the [C/Fe] abundance ratio that matches C$_24668$ best.  Once that
is known, it searches the [N/Fe] that matches CN indices best, then
it finally finds the [Ca/Fe] abundance ratio that matches the Ca4227
index best.  Note that the order in which those abundances are
performed is important, given that CN indices are sensitive to the
abundances of both carbon and nitrogen (as well as Fe), so that the
abundance of the latter can only be determined once that of the
former is known.  Models with the new abundances are then used to
redetermine age and [Fe/H] by returning to the H$\beta$$-$$\langle$Fe$\rangle$ pair,
in order to account for their small sensitivity to abundance ratios.
The process then is iterated until a final set of age and abundances
is converged to.  Convergence is typically achieved in no more than
a couple of iterations.  Readers interested in further details on
both the models and EZ\_Ages details on the models and age/abundance
determination methods, are referred to \cite{ezages} and
\cite{schiavon07}.  In this paper, we focus on the results for ages and
[Fe/H], deferring the discussion of abundance ratios to a forthcoming paper
(Schiavon et al. 2010, in preparation).

\subsection{Caveats}

Before discussing the ages and metallicities resulting from application
of EZ\_Ages to the data displayed in Figure~\ref{hbhdfem}, we
highlight a few of the limitations of our method.  First, EZ\_Ages
extrapolate from the models, and as a result,
ages and abundances cannot be obtained for clusters falling off the
model grids in Figure~\ref{hbhdfem}.  Therefore, clusters with
metallicities lower than [Fe/H]$ \sim -1.3$, or higher than [Fe/H]
$\sim +0.2$ are excluded from the analysis.  Also excluded are
clusters falling below the
oldest models, for which EZ\_Ages would find nonphysically old (i.e.,
older than the universe) solutions.  Most of these clusters are in
fact formally consistent with physically acceptable ages, given the
index error bars.  However, even after S/N considerations, a small
minority of the clusters are still consistent with older$-$than$-$the$-$universe
ages.  This may possibly result from a combination of model zero$-$point
uncertainties due to inadequate treatment of stellar luminosity
function and abundance ratio effects \cite[e.g.,][]{schiavon02} and
Balmer$-$line infill from evolved giants and/or intra$-$cluster medium
\citep{schiavon05}---see \cite{poole} for a discussion of this
issue.  Second, the treatment of horizontal branch stars in Padova
isochrones, adopted in the \cite{schiavon07} models, is 
such that there are blue horizontal branch stars in the metallicity
range spanned by those models.  As a result,
metal$-$poor clusters with a blue HB tend to have stronger Balmer
lines than predicted by models for the same age and metallicity,
with the result that their ages are artificially younger according
to the models.  This can be seen on the top panel of Figure~\ref{hbhdfem},
where the ages of clusters with [Fe/H]$ \simless -1.0$ are on average
apparently $\sim$ 8 Gyr or younger.  This is a well$-$known problem that has
been discussed in detail in previous works \cite[e.g.,][]{pacheco,
lee00}, and which at present has no satisfactory solution,
in view of the absence of a physically motivated theory capable of
predicting the envelope mass (and consequently the effective
temperature) of a star of given initial mass and chemical composition,
when it reaches the core$-$He burning phase.  Stellar population
models relying on
prescriptions based on free parameters have limited predictive
power for ages.
Our approach to this problem is that of singling out systems
where the presence of blue HB stars may render spectroscopic age
determinations unreliable.  A method has been developed for that
purpose by \cite{schiavon04}, which relies on the differential
effect of blue HB stars on higher and lower order Balmer lines
(H$\delta$ and H$\beta$, respectively).  Because blue HB stars are
brighter in the blue they affect H$\delta_F$$-$based ages more strongly
than those based on H$\beta$, which can be promptly seen from
comparison of the top and bottom panels of Figure~\ref{hbhdfem}.
Metal$-$poor GCs in the bottom panel appear to have ``younger'' ages
than on the top panel, suggesting the presence of blue HB stars in
these clusters.  It is therefore not surprising that there is a
significant difference between H$\beta$ and H$\delta_F$$-$based ages
for clusters more metal$-$poor than [Fe/H]$ \sim -0.95$ in the bottom
panel of Figure~\ref{age_feh}.  For these metal$-$poor clusters, ages from
H$\beta$ and H$\delta_F$ differ by 2.7 $\pm$ 1.4 Gyr, and we deem their
ages and abundances unreliable, excluding them from the present analysis.

\subsection{EZ\_Ages Results}

\begin{figure}
\vspace{1.0cm}
\plotone{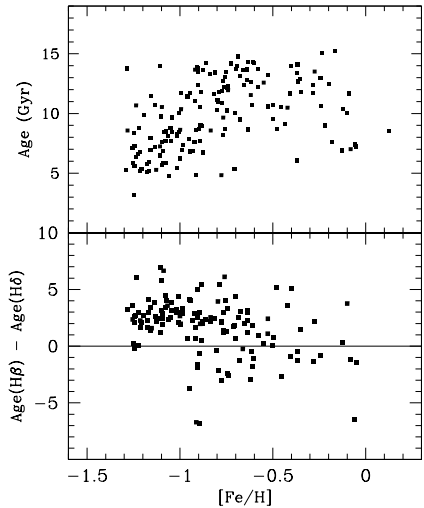}
\caption{ 
Top Panel: Ages and iron abundances resulting from
application of EZ\_Ages to the data shown in Figure~\ref{hbhdfem}. 
Bottom Panel: Difference between ages determined from $H\beta$ and
$H\delta_F$, as a function of iron abundance.  Ages for clusters more
metal-poor than [Fe/H]$ \sim -0.95 $ can not be trusted due to the presence
of blue horizontal branch stars, which are not considered in the
\cite{schiavon07} models, leading to artificially younger ages for
metal-poor systems.  The same effect is responsible to the systematic
difference between the H$\beta$ and H$\delta_F$ ages for metal-poor systems
in the bottom panel.  Ages for clusters with [Fe/H] $ \simgreater -0.95 $ are
reliable.  See text for details.
 \label{age_feh}
}  
\end{figure}
Results from the application of EZ\_Ages are shown in Figure~\ref{age_feh}.
In the top panel, [Fe/H] is plotted against spectroscopic age for
all the clusters falling within the model grids in the top panel
of Figure~\ref{hbhdfem}.  Following the above discussion, we do not
consider clusters with [Fe/H]$ \simless -0.95$.  The higher 
metallicity group occupies a locus in the age$-$metallicity
space characterized by a uniformly old population, with an age of
11.8 $\pm$ 1.9 Gyr, ranging in metallicity from [Fe/H]$ \sim -0.95$
to solar.  Because older clusters may have been left out of the
analysis, and because of the model and data limitations discussed
above, we make no claims on the absolute age of the oldest globular
clusters in M31.  On the other hand, EZ\_Ages can provide very
accurate relative ages (the internal age uncertainties are 2 Gyr), 
and with that in mind, we call attention
to the fact that the mean ages of the old GCs in M31 seem to be
remarkably constant over about a decade in metallicity ($-0.95
\simless $ [Fe/H]$ \simless 0.0$).  
The apparent, small trend of age with metallicity could be artificially
generated by a
dependence of HB morphology with metallicity that is slightly different from
that contained in the Tololo spectral sample, which is relatively devoid of clusters
with blue HBs with $-0.95 <$[Fe/H]$< -0.7$.  This small
trend will be examined in a future paper. 
One is of course also left wondering whether age remains
nearly constant towards the lower metallicity regime, where age determination
is unfortunately rendered unreliable due to the lingering gaps in
our knowledge of stellar evolution in the post$-$He flash phase, as
discussed above.

The ages found here are also recorded in Table \ref{main}. 
However, the
[Fe/H] values listed are those from Section \ref{s:emp_feh} above, for reasons
already discussed.  If the
indices for the cluster fell outside of the grids of Figure~\ref{hbhdfem},
or if the metallicity was lower than -0.95 as determined from the method of
Section \ref{s:emp_feh},
the age was set to 14.
\begin{figure}
\vspace{2.0cm}
\plotone{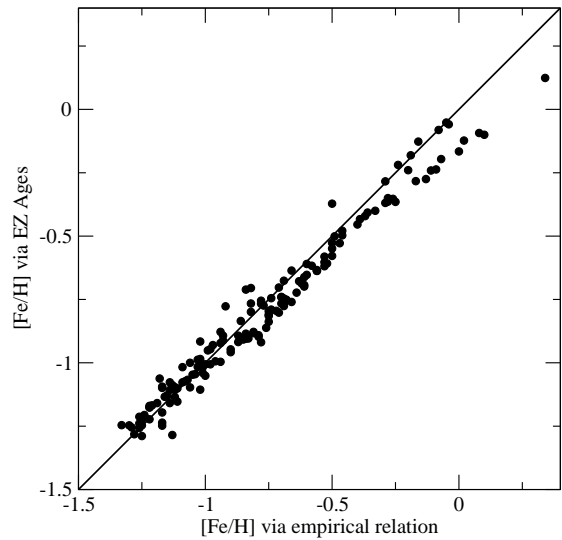}
\caption{[Fe/H] as derived by empirical fitting of galactic
GC indices compared with [Fe/H] from EZ\_Ages. The line represents equal values in
the axes.  For  [Fe/H]$>-0.95$, the mean offset is small, between 0.05 and 
0.08dex. \label{feh_comparison}}
\end{figure}
Figure \ref{feh_comparison} shows a comparison of [Fe/H] measured from Section \ref{s:emp_feh} and
the values using EZ\_Ages. This indicates excellent agreement between the two methods 
for [Fe/H]$>  -0.95$, with a small mean offset of 0.05$-$0.08dex.
The detailed discussion of the M31 GC metallicity distribution will be taken up
in Section \ref{s:metallicity}. First, we compare the [Fe/H] values
derived from the MW GC calibration
with those from other sources.

\section{Comparison of [Fe/H] with other measurements}

\subsection{[Fe/H] from other Integrated Light Studies}
\begin{figure}
\vspace{1.0cm}
\plotone{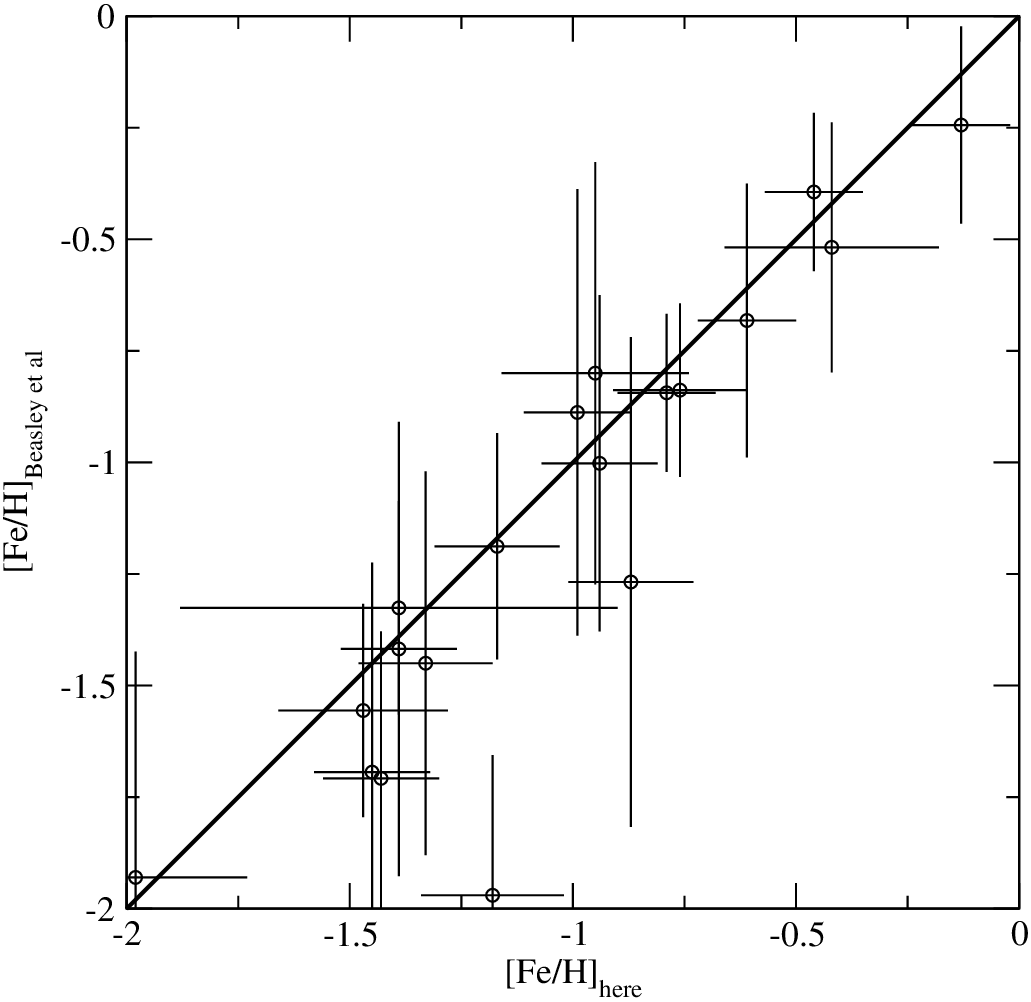}
\caption{Comparison of [Fe/H] derived from Hectospec spectra and inferred [Fe/H] values from
\cite{beasley}, derived from spectra also taken with the MMT, but using the
Blue Channel spectrograph. The stellar population models  used for the 
Beasley results plotted here were those of \cite{TMB}. [Fe/H] values were calculated by
using the published [Z/H] and [$\alpha$/Fe] values.
The diagonal line represents equal values in
the axes. \label{beasleyZ}}
\end{figure}

\cite{beasley} derived [Z/H] values from MMT Blue Channel spectrograph
spectra for 23 M31 objects. One of these is a
star as noted below, otherwise we have 20 objects in common. Figure \ref{beasleyZ}
shows a comparison of our [Fe/H] values derived from the MW GC Lick index calibration, 
and the Beasley et al. [Z/H] values converted
to [Fe/H] using their published  [Z/H] and [$\alpha$/Fe] values (and assuming that
[Z/H]= [Fe/H] + 0.94[$\alpha$/Fe], from \citep{TMB}. (Beasley et al.
used two models to determine [Z/H] $-$ the differences between the two are
small.)  We find excellent agreement for the objects in common.

\cite{colucci}  measured [Fe/H], among other abundance ratios, for 5 M31 GCs
using high resolution spectra and a method that matches observed equivalent widths
of weak metallic lines with simple stellar population modelled equivalent widths.
Four of their clusters are in common with our measurements (B045-G108, B381-G315,
B386-G322, and B405-G351), and though the metallicity
range is limited ($-0.86$ to $-1.22$), the mean offset is 
gratifyingly small, 0.08 dex,
and the standard deviation of the differences is very small, 0.02 dex.

\subsection{[Fe/H] from New Color$-$Magnitude Diagrams}

\begin{figure}
\vspace{1.0cm}
\plotone{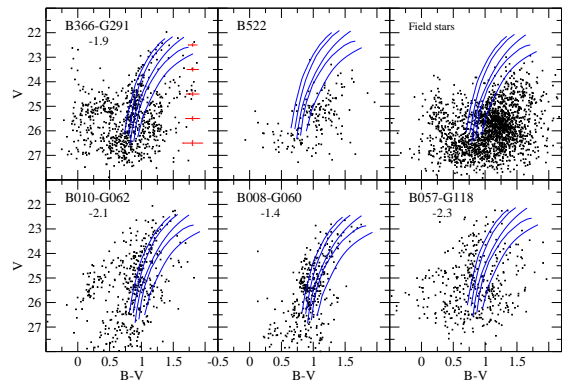}
\caption{CMDs of 5 old clusters and the field  for
B057-G118. Also shown are fiducial giant branches for 5 galactic
globulars with a range of abundances :  NGC4590 ([Fe/H]$=-2.3$), NGC 6809 ($-1.9$), 
NGC6752 ($-1.6$), NGC362 ($-1.3$)  and 47 Tuc ($-0.8$) \citep[G.S. Da Costa,
priv.comm., metallicities from][]{carretta09a}. 
Reddenings for the clusters listed in Table \ref{main} 
were used to shift the fiducial curves. A distance modulus of 24.43 was
assumed.  Abundances of the M31 clusters as derived from these CMDs are shown, 
and are listed in Table \ref{compare_feh} along with those derived from 
the spectra. Median photometric error bars are shown in the figure for 
B366-G291, at a range of magnitudes.  The field stars were collected over
an area 25 times larger than that for the clusters.\label{cmd}}
\end{figure}

We are fortunate that M31 is near enough to allow metallicity estimates to also
be made using the mean colors of giant branch stars as resolved in {\it HST} imaging. 
Thus comparisons between [Fe/H] for clusters 
derived from spectra and  CMDs  can be performed.
The HST/ACS images of the seven fields in the disk of M31 that were used to study
the CMDs  of several young clusters in Paper I also contain
more than 10 old clusters.  To add to the small but important 
literature of M31 CMDs,  we have obtained point$-$spread function (PSF) 
photometry of five old clusters
from that new data set (the other clusters are in extremely crowded fields). Data for four of 
these clusters were also analyzed by
\cite{perina}.
These {\it HST} data were processed  as described in Paper I.  In brief,
we used the DAOPHOT package of
\citet{1987PASP...99..191S}, modeling the spatially variable PSFs 
for each of the  combined images separately, using only 
stars on those images.  PSFs were constructed using $5-10$ bright stars which had no pixels
above a level of 20,000 counts, the point
at which an ostensible non$-$linearity sets in. 
Aperture corrections were also measured using these stars,
to determine any photometric offset between the PSF photometry and
the aperture magnitude within 0.5\arcsec.   \citet{2005PASP..117.1049S} have
provided aperture corrections from that aperture size to infinity, 
in all Advanced Camera for Surveys (ACS) filters. The photometry was then placed on the standard Johnson/Kron-Cousins
VI system using the aperture corrections and synthetic transformations 
provided in \citet{2005PASP..117.1049S}. To lessen the severe problems
with crowding in these clusters, only stars that fall in an annulus with
radii  of 15 and 50 pixels (0.75 and 2.5\arcs) from the center are shown in the
color$-$magnitude diagrams (Figure \ref{cmd}).

Figure \ref{cmd} shows the CMDs of the five clusters, along
with fiducial giant branches of galactic GCs, shifted assuming a distance modulus
of 24.43 and the clusters' reddenings derived below and listed in Table \ref{main}.
[Fe/H] values for the clusters were then visually estimated by the position of the observed
giant branch compared to the fiducial branches, and the
values are listed in Table \ref{compare_feh}.  The uncertainties in the derived
[Fe/H] values are 0.2 dex. A CMD comprised of field stars in
an area 25 times larger than the cluster areas is also shown.
B522 was too poor in giant
stars for an estimate to be made.  The giant branch of B057-G118 is bluer than our
most metal$-$poor fiducial cluster, but we have left the derived metallicity at
the minimum of $-2.3$.
\begin{figure}
\vspace{1.0cm}
\plotone{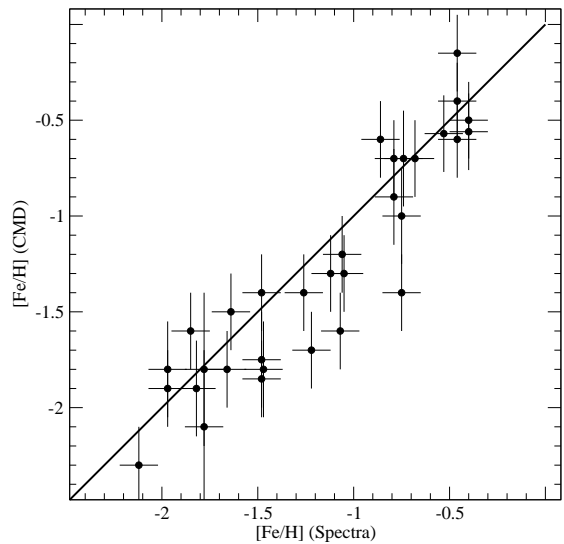}
\caption{Comparison of [Fe/H] derived from HST CMDs and from the spectra reported here.  Clusters
with more than one CMD value are shown multiple times, at the same value for [Fe/H](Spectra). 
The diagonal line represents equal values in
the axes.  The rms of the differences between the two methods is 0.2 dex. \label{cmdspectra}}
\end{figure}

Figure \ref{cmdspectra} shows the comparison of [Fe/H] values derived from our spectra and from
CMDs either from this paper, or from the literature, for 22 clusters.  These values are also listed in 
Table \ref{compare_feh}.
Errors in the spectroscopic values are all
0.1 dex, while most of the errors in the CMD values are around 0.2
dex, except for
the most metal$-$poor clusters for which we estimate the uncertainties
to be 0.4 dex because
the giant branch colors change slowly  with metallicity. Overall the agreement is 
surprisingly good, given
the dependence on assumed reddening  for the CMD values and the uncertainties inherent in the
spectroscopic values.  The tendency for the middle range metallicity clusters to have
higher spectroscopic estimates than those from CMDs could be due to small calibration problems with
one or both of the methods
The rms of the two methods is 0.2 dex.

\section{Cluster Masses}
Before we look further at the age and metallicity distributions, we must
attend to the detail of estimating cluster masses. 

\subsection{New Reddenings}
To estimate masses for the old clusters 
from the photometry in Paper I and assumed $M/L$ ratios, reddenings must be also known. 
Paper I described the
technique used here. To repeat in  brief, first the flux$-$calibrated spectra of clusters
with low$-$reddening were dereddened using the \cite{barmby} values.  
Then the spectra were ordered in metallicity, and 
interpolation formulae were created via a least$-$squares fit
for intensity as a function of both wavelength and
metallicity. This allowed a cluster spectrum of arbitrary
metallicity to be formed, dereddened to the accuracy of the \cite{barmby} reddenings. 
Reddenings for each cluster could then be found by comparing the spectra
with the metallicity$-$appropriate interpolated spectrum, 
and adjusting reddenings as needed to bring continuum shapes
close to that of the expected shape.  We assumed $R_{\rm V}$=3.1, though
there have been some reports that  $R_{\rm V}$ is lower in M31
\citep{iye, sharov}.
Thus this method is similar
to methods that used relations of metallicity and color among MW GCs
to derive reddening$-$free colors of M31 clusters which had spectroscopic 
metallicity estimates (as was done by Barmby et al.), except that we used our flux$-$calibrated
spectra to derive both the metallicities and colors.

The interpolation formulae were used 
to derive reddenings for the 150 GCs for which we have
spectra and whose reddenings were not measured in \cite{barmby}, 
as well as to modify the reddenings that had been estimated by
Barmby et al. 
Where we could not determine
new reddenings because the spectra could not be accurately fluxed
(some observations were taken with a malfunctioning ADC), we use
the modal reddening of $E(B-V)$=0.13. Table \ref{main} lists the 
reddenings for the clusters.
\begin{figure}
\vspace{1.0cm}
\plotone{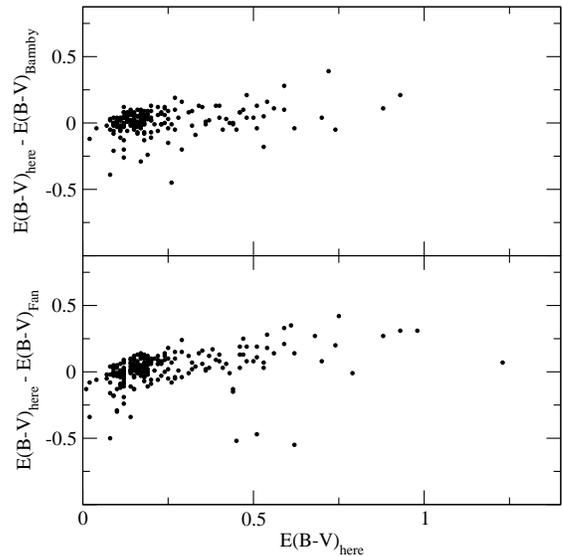}
\caption{Comparison with \cite{barmby} and \cite{fan} reddenings.
The differences in reddenings determined here and by \cite{barmby}, top 
and \cite{fan}, bottom, are plotted against the reddenings given here in 
Table \ref{main}.
 \label{compare_red}}
\end{figure}

Figure \ref{compare_red} shows a comparison of our $E(B-V)$ values with 
those for the same clusters contained in 
\cite{barmby} and \cite{fan}.  Since the \cite{barmby} 
reddenings were used as the starting point, it is not surprising
that there is no mean offset, but it is gratifying that the rms is
small, at 0.09mag.  The mode of our $E(B-V)$ values, like that
of Barmby et al., is  0.13. The comparison with \cite{fan} is good, 
except for 10 clusters where they show reddenings larger than ours  by $>$ 0.5mag.  
These differences are not correlated with any obvious parameter
such as reddening value or metallicity, so
we have no explanation.  Our reddenings are about
0.1 smaller than those of \cite{fan} at low reddenings, and about 0.2 mag redder at reddenings around 0.6.
\cite{vdb2} reported a mean M31 foreground reddening of 0.06,  and 
\cite{massey} reported a mean reddening of 0.13 for OB associations, in fair and good
accord respectively with the mode of our values.  We estimate the uncertainties in
our reddenings to be 0.1 mag.

\begin{figure}
\vspace{1.0cm}
\plotone{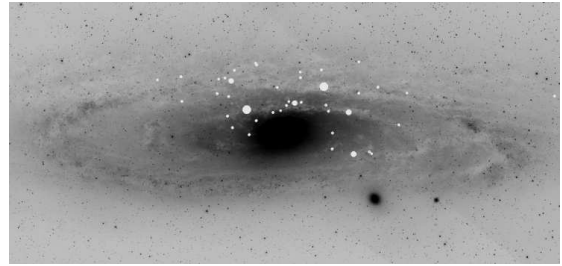}
\caption{Location of highly reddened clusters (image from the DSS, the
diameter of the disk shown is 1.8\degr = 26 kpc). All white circles have
reddening $E(B-V)>0.4$. Largest circles have  $E(B-V)>1.2$; intermediate
sized circles have reddening between 0.8 and 1.2. \label{red_dist}}
\end{figure}

Figure \ref{red_dist} shows the location of the most highly extincted clusters. As noted
by \cite{fan}, these are mostly found on the NW side, which is the nearer side (the upper
part in this rotated figure). B037-V327 still has the highest reddening known in M31 at $E(B-V)$=1.6, followed 
closely by B129 with 1.2.  The former cluster actually has differential reddening across its
face \citep{ma}, and this is confirmed in our analysis,  by the exceptionally 
poor correction to an unreddened spectrum. That is, using a single reddening value for
the spectrum does not result in a spectrum with a continuum shape
that matches other clusters with its metallicity, or
any metallicity. No such problem is seen for any other cluster, thus we would state that no
other cluster has significant differential reddening (though the {\it HST}
image of B151-G205 shows some strong extinction in its outer parts).  The reddening
for  B037-V327 is higher than found by \cite{barmby} (1.4) and much higher than found by
\cite{strader3} (0.92), who forced the cluster to fall on their GC fundamental plane. 
Our high value results in the cluster becoming the most massive cluster associated
with M31, by a factor of 2.5 over the next cluster (B023-G078). Using the Barmby value
still leaves it as the most massive, but brings it more in line with the other
clusters.

There are 13 known MW clusters with reddenings higher than any of the known M31 clusters, which
can be understood as being due to our vantage point in studying M31, not looking through its disk
plane (the highest extinction MW clusters are seen at low galactic latitude).  Still, it is
possible that some M31 clusters remain to be found that are currently hidden behind dust lanes
in the disk of M31.

\subsection{Calculating the Cluster Masses}

To calculate the photometric masses of the clusters, 
we assumed $M/L_{\rm V}$=2 independent of [Fe/H] .
Stellar population models have long predicted that  $M/L_{\rm V}$
should increase with  [Fe/H] if  age and initial mass function are fixed \citep{tinsley}, 
but recently  \cite{strader3} have shown for a small number of M31 GCs that
$M/L_{\rm V}$ apparently declines with [Fe/H].  Since the matter is
an active area of study, we elected to use a constant at this time and chose
the value of 2 from examining the values measured in \cite{strader3} .
The $V$ band photometry
was taken from Table 1 of Paper I
\citep[obtained from images reported in][]{massey}, and the reddenings 
from Table \ref{main} of this
paper.

Before we discuss the results, we should comment on the completeness of
our catalog of M31 GCs. While we have provided much new data on
the clusters contained in the catalog (including whether the catalog
entry was in fact a GC),
the construction of the input catalog itself 
was not systematic, being more of a historical document \citep[see][]{PaperI, galleti2}. 
It is likely to be nearly complete at the bright end,
but the completeness at the faint end is unknown.  Indeed, as we write, a new
survey of the disk of M31 is being undertaken which may find many more old but
low$-$mass clusters
projected on the disk ({\it HST} program GO 12055, PI Dalcanton).  
Thus, our comments about the mass distribution here will
be limited.
\begin{figure}
\vspace{1.0cm}
\plotone{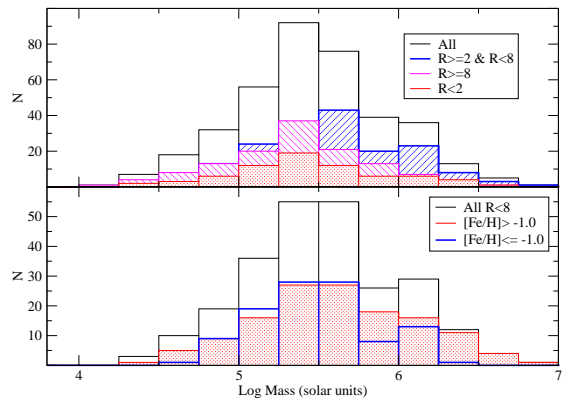}
\caption{Histogram of total masses of M31 GCs. Top: Clusters divided into three radial bins. Clusters
inside of 8 kpc but outside of the bulge (R$>2$ kpc) tend to be more massive than those in the other
two radial bins. Bottom: Clusters with R$<8$ kpc divided into two
bins of metallicity. There is no obvious difference in the two mass distributions, 
aside of a tendency for the most and least massive clusters to be more
metal-rich than [Fe/H]=$-1$. \label{masses_vs_R}}
\end{figure}

Figure \ref{masses_vs_R} shows the distribution
of derived GC masses, divided into radial and metallicity bins, 
indicating
that the masses extend from about $2 \times 10^4$ to $10^7  M_\sun$, with 
the most massive cluster being B037-V327 in our calculations (but see the caveat in
the previous section).  It is interesting that
our six most massive
clusters all have reddenings greater than 0.4 mag.  The median cluster mass 
is about $3 \times 10^5  M_\sun$ (or $M_{\rm V}=-8.0$), which is about twice
that of the MW median cluster \citep[for which $M_{\rm V}=-7.4$,][]{harris}.
There may be a difference in the distributions when separated into mass bins, 
in the sense that more massive clusters tend to prefer
the projected area outside of the bulge area ($R$=2 kpc) but within 8 kpc.  This could be due
to the most massive clusters near the bulge being destroyed. If the clusters within 8 kpc are 
divided in two parts at a metallicity of [Fe/H]=$-1.0$, we see another slight tendency for the
metal$-$rich group to have both more massive and less massive clusters.

\begin{figure}
\vspace{1.0cm}
\plotone{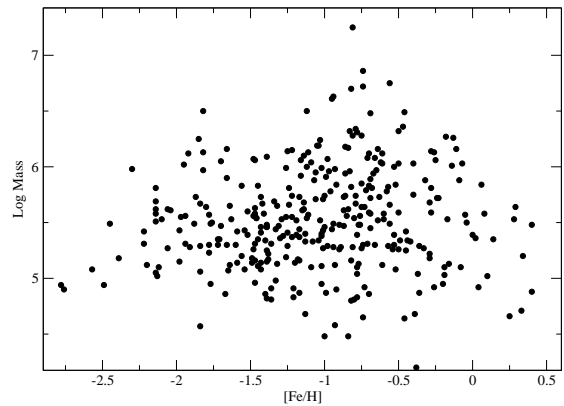}
\caption{Distribution of M31 clusters in the metallicity-mass plane. Mass is shown
in log of solar masses.  There is no evident trend of mass and metallicity for
low metallicity clusters as might be expected if there was a ``blue tilt'' \citep{strader2}.  
Thus, there is no evidence for self-enrichment in the metal-poor clusters in M31.
\label{mass_metal}}
\end{figure}

Figure \ref{mass_metal} shows there is no
evident relation of mass and metallicity for low metallicity clusters,
which might have indicated self$-$enrichment in the metal$-$poor clusters  (the ``blue$-$tilt'').
This is in accord with Figure 21 of the \cite{barmby} M31 paper.
\cite{strader2} noted that such
a relation is not seen for MW clusters either, though it is common in
early type galaxies \citep{harris09}.
This figure also demonstrates that
the most metal$-$rich clusters are not the most massive.

\section{Discussion of Ages}
\label{s:ages}
We now discuss the issue of age variations among the M31 globular clusters, comparing
the results found using EZ\_Ages and the statements of previous authors that a number
of M31 clusters have intermediate ages.

\begin{figure}
\vspace{1.0cm}
\plotone{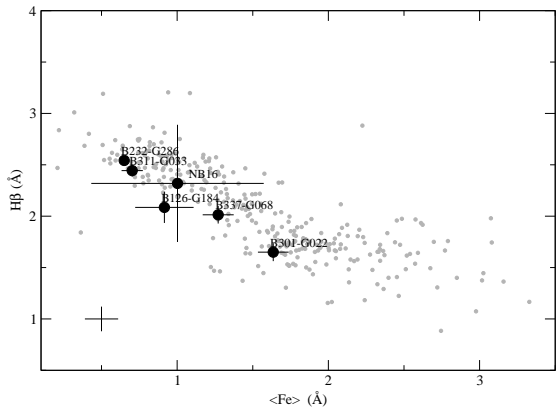}
\caption{$\langle$Fe$\rangle$ (average of Lick indices Fe5270 and Fe5335) plotted against  H$\beta$. All
of the old clusters are plotted in small, gray filled circles. The six clusters previously reported
to be intermediate age are shown as large black circles, with their names attached.  None of
these shows any evidence for enhanced H$\beta$ strength, and hence none has evidence for
an intermediate age.   Average error bars for all the clusters are
again shown in the bottom left corner.
 \label{notyoung}}
\end{figure}

Based on archival spectra,  \cite{beasley} stated that, in particular, B158-G213 and B337-G068 had
similar metallicities but different ages.  They also stated that
five M31 clusters were intermediate in age, with ages between 2 and 5Gyr: 
B126-G184, B301-G022, NB16, NB67 and
also, B337-G068.  \cite{brodie} compared the spectrum  of NB67 with two other M31 clusters 
and concluded that it was an intermediate age cluster.
Our response to those claims follows.
From the Hectospec data in Table 
\ref{main}, we found that (a) B158-G213 and B337-G068 are old and 
have dissimilar metallicities
($-0.8$ and $-1.2$, respectively),
(b) B126-G184, B301-G022, NB16, and  B337-G068 are all older than 9 Gyr, with [Fe/H]  values
ranging from $-0.8$ to $-1.5$. 
As it turns out, NB67 is a foreground F star \citep{PaperI}.
Using at the time fresh data, \cite{burstein2} additionally claimed that B232-G286 and B311-G033 had ages of 5 Gyr.
We found that both of these clusters are old, and simply very metal$-$poor ( [Fe/H] = $-2.0$ and $-1.9$, respectively).
In general, previous authors have mistaken lower metallicity clusters for younger ones.  To demonstrate our
claim graphically, Figure \ref{notyoung} 
highlights the six purportedly intermediate age clusters in the $\langle$Fe$\rangle$$-$H$\beta $ index diagram, similar to 
the top panel of Figure \ref{hbhdfem}.  If these clusters were substantially younger than the mean cluster
age at any given metallicity (as measured by  $\langle$Fe$\rangle$), we would expect them to have H$\beta $ indices 
stronger than the mean  H$\beta $ index.  Such is not the case.

A series of papers using the BATC photometric system combined with other photometry
has produced several
tables of ages for clusters, young and old \citep{fan2, ma2, wang}. The ages were 
derived using SSP models, based on Padova isochrones. There is very
little correlation of the ages in those papers and those we have reported here and in Paper I.
The main source of the discrepancy is that many of the clusters identified by the cited papers as
being young are in fact old and metal$-$poor.  Of the 77 clusters in common with \cite{wang}, 50
clusters stated to be younger than 5 Gyr are older than 10 Gyr based on our analysis.

Returning to our own age determinations, we note again that from the EZ\_Ages analysis, 
some clusters marked as old in Paper I 
were realized to be
younger than 2 Gyr after publication of that paper. 
These are all disk clusters; Table \ref{revised} lists these
clusters with their revised ages, all under 2 Gyr, as derived from the method of Paper I
and confirmed by EZ\_Ages.
Aside of those, there are a small number of clusters (12) with ages younger than 8 and older than 2 Gyr, but
six of these have abundances close to the problem [Fe/H] value of $-0.95$ and whose
ages are thus suspect (see above). Thus
only six are worth further consideration with regard to intermediate ages. These are
B015, B071, B138, B140, B268, and AU010. Their ages are all around 7 Gyr, 
and all but B015 are within 2 kpc of M31's center.  These have masses between
 $10^5$ and $4 \times 10^5 M_\sun$, close to the median mass for all the M31
GCs.  Interestingly, they are all metal$-$rich (five out of the six have [Fe/H] $> -0.2$,
and have very strong CN bands.

\begin{figure}
\vspace{1.0cm}
\plotone{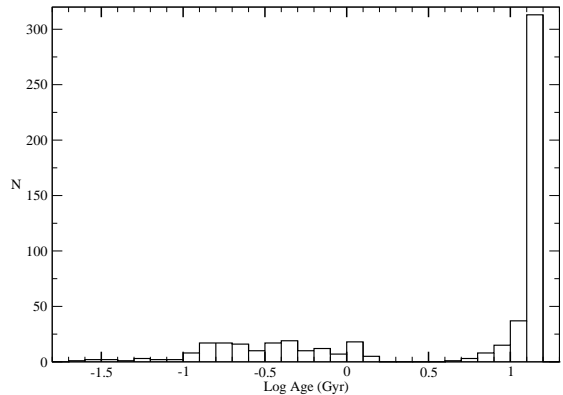}
\caption{Histogram of ages. Ages for clusters older than 1 Gyr were
determined via EZ\_Ages, ages for younger clusters come from \cite{PaperI}. Ages for old
clusters with [Fe/H]$<-0.95$ were set to 14 (1.15 in the log).  \label{age_hist}}
\end{figure}
To conclude the age discussion, we find no evidence for any massive clusters in M31
with intermediate ages, those between 2 and 6 Gyr.
Figure \ref{age_hist} shows the age histogram, including the young clusters whose
ages were determined in Paper I. For this diagram, we assumed that all clusters with 
[Fe/H] $< -0.95$ (whether we determined the metallicity here or others did so
elsewhere) have ages of 14 Gyr. We also required the clusters to have  masses
greater than $5 \times 10^3  M_\sun$. This diagram clearly shows the gap in ages
between  2 and 6 Gyr.
Moreover, we have found
that the mean ages of the old GCs in M31 seem to be
remarkably constant over about a decade in metallicity ($-0.95
\simless $ [Fe/H]$ \simless 0.0$).

\section{M31 GC Metallicity Distribution}\label{s:metallicity}

Next, we discuss the distribution of metal abundances 
for the entire M31 GC sample. For clusters we did not observe
with Hectospec, we have
drawn [Fe/H] values from the literature, but only when
[Fe/H] values were derived from
CMDs. \cite{huxor} supplied 14 values for their distant sample, and 4 others
came from other sources. These are G001, B468, B358-G219 \citep{fusi}, 
and B293-G011 \citep{rich}.  By including these other clusters, we have
expanded the discussion radius out to 100 kpc, though unfortunately many of the
distant clusters do not yet have published, accurate  [Fe/H] values (25 out of
38).
\begin{figure}
\vspace{2.0cm}
\plotone{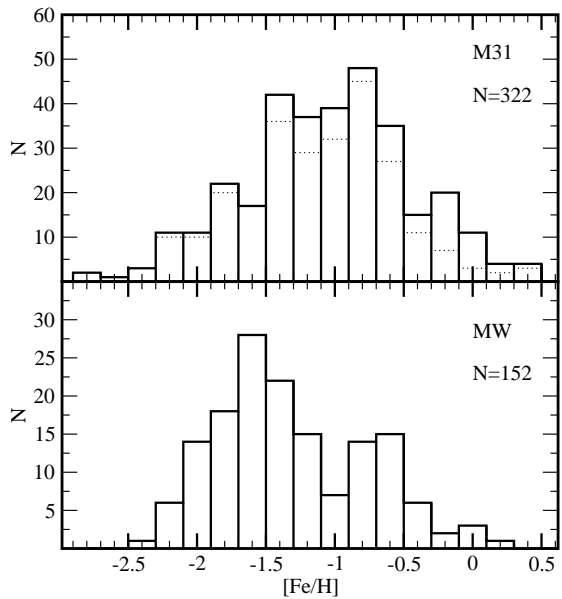}
\caption{Comparison of metallicity distributions for 322 M31 globulars 
mostly derived from spectra here (top
  panel), and for 152 Milky Way globulars, for which 
the metallicity sources are described
in the text (bottom panel).  The dotted histogram in the M31 panel
shows that the distribution changes very little if clusters with
a distance of less than 2 kpc from M31's center are excluded.  \label{z_dist}}
\end{figure}
Figure \ref{z_dist} shows the distribution in [Fe/H] for the 322 M31
clusters where the estimated error in [Fe/H]
is less than 0.5 dex or the spectral S/N is greater than 20. We do not
show clusters associated with NGC~205 here, since they are still bound to that
galaxy.
The distribution is not generally bimodal, because of the lack of
a strong minimum, rather it shows a broad peak, centered at  [Fe/H]=$-1$ . 
A Kolmogorov$-$Smirnoff test indicates that the observed distribution has
a 28\% chance of being drawn from a single Gaussian distribution, with
mean  [Fe/H]=$-1.08$ and sigma=0.61. (By contrast, the MW distribution
has essentially no chance of coming from a single Gaussian.)  To be clear,
we are not stating that the distribution is unimodal, rather we find the
distribution not to be clearly bimodal.
Visually, the observed
distribution could also contain two equal sized 
Gaussian distributions centered at  [Fe/H]=$-0.7$ and $-1.4$, for instance,
and possibly a third peak at [Fe/H]=$-0.2$, though such interpretations
can not be verified statistically. A mixture$-$model
KMM test \citep{ashman2} of two and three peaks,
with either homo$-$ or heteroscedastic variances reveals that such multimodes
are not statistically significant.
If we consider only the 257 clusters outside of the bulge area (roughly those
farther than 2 kpc from the center), the peaks at  [Fe/H]=$-0.7$ and $-1.4$
become visually slightly more prominent, though the KMM test still shows no
significance to them.
Also, even with that artificial restriction, there
is certainly no indication that the M31 metal$-$poor peak is similar 
in strength or in location
to the MW metal$-$poor peak at [Fe/H]$=-1.6$ (bottom panel of Figure  \ref{z_dist}).

The [Fe/H] distribution obtained from the Mg~b index 
does show a somewhat more
definite peak in the M31 clusters at $-1.4$ (Figure \ref{hist_compare}),  still more 
metal$-$rich than
the MW metal$-$poor peak. By further inspection, we found that 
all of the line indices that have strongly bimodal distributions
also have inflections in the index versus [Fe/H] relation at low
metallicity (between [Fe/H] $ -1.7$ and $-1.2$, varying from index to index).
The inflection is responsible for the piling up of
low$-$metallicity clusters at constant index values, which leads to the
low$-$metallicity peak in the index histograms (low values for metal indices,
high values for Balmer indices).  Therefore, the low$-$metallicity peak in
those index histograms does not translate into a corresponding low$-$metallicity peak,
but is rather a by$-$product of the index$-$[Fe/H]
relation, which is such that metal$-$poor clusters in a wide range of
metallicities have approximately the same index values.  

Whether a peak at [Fe/H]=$-1.4$ is real or not will require better
calibration than we currently have, in particular more metal$-$poor MW GCs are necessary. 
We maintain that our  $\langle$Fe$\rangle$$-$[Fe/H] method is the best
one to use $-$ previous calibrations used stronger lines, and even less data to calibrate
the MW GCs $-$ but the calibration here must still be seen as preliminary, and 
thus the M31 [Fe/H]  distribution
is not yet firmly established, though we are much closer than previously. 
The M31 GC metallicity  distribution is definitely different from the MW GC  distribution, and
certainly it is not simple to divide the clusters into two groups  as has been done for the MW GCs.
This fact must indicate that the formation of the M31 cluster system was substantially different
from that in the MW.

\subsection {Comparison with MW GCs}
In order to compare the metallicities of M31's old clusters
with the Milky Way's, we took the MW cluster data
originally  compiled by
\citet{harris}, and subsequently updated by \cite{bica06}, and
further updated it with 17 new metallicities of
\cite{carretta09a}.
This results in values for 152 MW clusters,
whose distribution is shown below that of the M31 clusters in Figure  \ref{z_dist},
indicating the well$-$known peaks at  [Fe/H]=$-1.6$ and $-0.6$ \citep[e.g.,][]{bica06}.

It  is interesting to note that
\citet{huchra} presented a metallicity distribution for the M31 globulars
which was quite similar to the Milky Way's, prompting the 
comment ``Like the Milky Way, only more so.''  We do not find this to be 
the case. The modal metallicity in
their M31 histograms was quite low, near [Fe/H]=$-1.7$. As more clusters
which were projected on the disk were studied, the modal metallicity
moved higher, to $-1.3$ \citep{barmby} and then to $-1.1$
\citep{perrett}. Both our measurement and that of \citet{galleti09}, based on
metallicities found in the literature prior to this paper, agree
on a modal metallicity that is actually higher than [Fe/H]=$-1.0$. 
\cite{ashman} derived a bimodal metallicity distribution
using the data of \cite{huchra}, and \cite{barmby} reported marginal
confirmation of  bimodality using $U-V$, $U-R$ and $V-K$ colors (marginal in the
sense that bimodality was found at the 92$-$95\% confidence level). \cite{fan}
used literature spectroscopic metallicity values and newly derived metallicities
from colors and also derived a bimodal M31 GC metallicity distribution (significant at
a level greater than 95\%). However, all of those studies suffered from the  inclusion
of clusters now thought to be young, which may explain in part why those studies
found more significance to bimodality. An additional factor must be the conversion
of colors to metallicity \citep{yoon},  which is non$-$linear at the metal$-$poor end, leading to
further difficulties interpreting bimodality.




It has been known for some time \citep[see, for
example][]{mouldkristian} that the field stars in M31's halo are
significantly more metal$-$rich than the MW halo's field
stars. This, combined with the early results on the M31 cluster
metallicity distribution being similar to the MW clusters,
prompted \citet{durrell94} to point out that there was a curious
offset between the M31 halo cluster and field metallicities.
\citet{freeman} suggested that perhaps the $R^{1/4}$ bulge
contributed significantly to the halo field (but not the globular
clusters) at large distances from its center.  However, it now seems
that the metallicity distribution of old clusters in M31 is also
relatively metal$-$rich, removing the need for such distinctions.

What of the extremes of the distribution? Do we have evidence that the
globulars in M31 have metallicities that reach higher or lower than
the MW system? The M31 cluster distribution does apparently extend
further to the metal$-$rich and metal$-$poor ends than the MW's, 
but for the metal$-$poor end the most we can say is that there are 
M31 clusters as metal$-$poor as those in the MW, such as NGC7089 (which
has [Fe/H]=$-2.4$).  Our technique of using only Fe indices to measure
[Fe/H] has a natural failing at the metal$-$poor end because of 
the decreasing S/N of those features, which is compounded by
the break in the relation of the Lick Fe indices and [Fe/H] for
MW clusters (see Figure \ref{MW indices}).  The clusters B157-G212 and B028-G088
have very high S/N spectra and are definitely as metal$-$poor
as NGC7089, but whether they are more metal$-$poor as our table indicates 
would require higher
spectral resolution data or individual star spectra.

However, there are several M31 old clusters
with [Fe/H] greater than solar and low measurement errors. Such
super$-$metal$-$rich globulars are not seen in the MW. While
previously it was possible to suppose that these clusters existed in
the MW but were hidden by dust extinction, both the 2MASS and
GLIMPSE surveys have only identified a handful of new clusters. 
This means that our catalog of MW GCs is close to
complete. It is likely that we are seeing a real difference in the
metallicity distributions at the metal$-$rich end, perhaps caused by the
significantly larger number of M31 old clusters.  By contrast, the
absence of strong evidence for more metal$-$poor clusters in M31
\citep[added to the dearth of clusters with metallicity less than
$-$2.5 in the LMC, Sgr and Fornax dwarfs,][]{mackeygil04} hints that
there is some physical mechanism which prevents the formation of
extremely metal$-$poor globular clusters.

The other obvious difference between the metallicity distributions of
M31 and MW GCs is the bimodality in the Milky Way
distribution, which has been known for some
time.
In an influential early paper,
\citet{zinn2} pointed out the clear division of the globular clusters 
into two systems, which he named the halo and disk
clusters. 
This division can be seen clearly in a number of different
properties of the clusters. First, MW disk clusters have small distances
  from both the galactic center and the galactic plane while halo clusters
  extend out to large values of both.
Second, the MW disk GCs are on the metal$-$rich side of the metallicity
distribution, having [Fe/H]$>-0.8$, and the halo clusters are of course
metal$-$poor. Finally, the disk globulars are rotationally supported,
  although they have hotter kinematics than the Galaxy's thin disk,
  while the halo globulars have a mean rotation close to zero and a
  high velocity dispersion.

\cite{minniti} and \cite{cote} later suggested that
the metal$-$rich clusters
are associated with the galactic bulge or perhaps the bar.
Others have
proposed divisions into young halo, old halo, bulge/disk \citep{mackey, fraix}
based on numerous observed properties of the clusters.

While the M31 globulars do not have a clearly bimodal metallicity
distribution, it is interesting to examine the variation of
metallicity with position in M31. Because we do not have 3$-$D positions
of our clusters in M31, we cannot simply plot [Fe/H] against $z$
distance to make comparisons with Zinn's results; but because M31 is
quite highly inclined to the line of sight, we can reasonably use the
distance from the major axis, Y, as a proxy.

\begin{figure}
\vspace{1.0cm}
\plotone{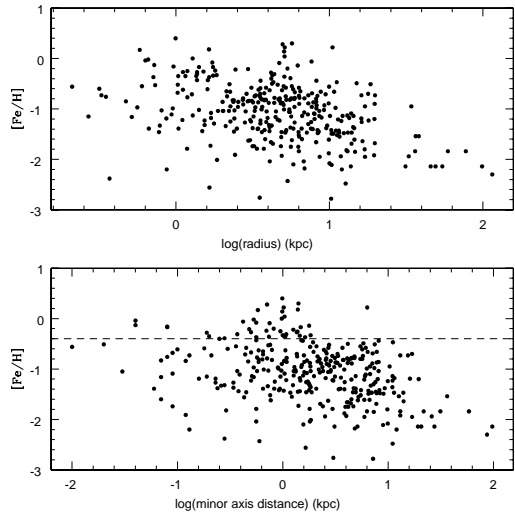}
\caption{Relation between metallicity and position in M31. In the top
panel we show the variation of [Fe/H] with log(radius), and in the bottom
panel with the log of the minor axis distance Y, which is the
closest comparison we can make to the 3-D distance above or below
the galactic mid-plane.  The dotted line in the bottom panel
marks [Fe/H]= $-0.4$. Clusters with higher metallicities than this appear
to be confined to distances close to the midplane.
\label{radial_y_fe}}
\end{figure}

We can contrast the spatial and chemical properties of the two old
cluster systems by examining Figure \ref{radial_y_fe}, which plots
both projected radial distance from M31's center and the minor axis
distance Y against [Fe/H]. We see that M31 shows marked similarities
to the MW populations first noted by Zinn, with some intriguing
differences. We see that, as in the MW, the most metal$-$rich
clusters in M31 are restricted to its inner regions, both in radius
and in minor axis distance, which is our best proxy for height above
the plane. Conversely, the more metal$-$poor clusters occupy a large
range in radial and minor$-$axis distance. 
However, in M31 it is only the clusters with [Fe/H] greater 
than $-0.4$ (see the dotted line in the bottom panel) which are confined 
to distances close to the midplane. M31 clusters with [Fe/H] 
greater than $-0.4$ have a median $|$Y$|$ distance of 1 kpc, while 
clusters with [Fe/H] less than $-0.4$ have a median $|$Y$|$ distance 
of 2 kpc.
We also see a hint of a new feature: a third
break around [Fe/H]=$-1.5$. The outer halo clusters, which reach out to
$\sim$100 kpc in projection, all have metallicities lower than
$-1.5$. This is a numerically small group even in M31, and it is
possible that the smaller total number of MW GCs has
obscured a similar pattern. Spectroscopic metallicity measurements of
these M31 far outer halo globulars will also be useful for more precise
comparisons.

\subsection {Radial Distribution of Clusters}

\begin{figure}
\vspace{0.0cm}
\plotone{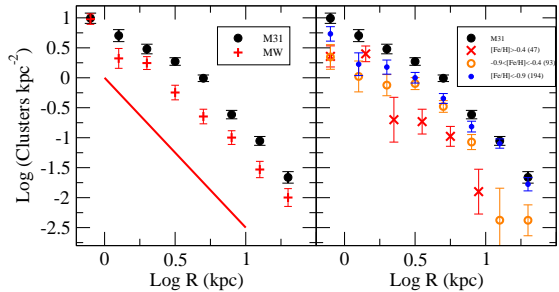}
\caption{Left:Radial profile of the M31 cluster system compared to that for MW clusters.
Projection onto the YZ plane was used for the MW clusters.  Circular symmetry was assumed
for both galaxies. The solid line represents a
$-2.5$ power law, which corresponds to the projected MW  halo profile. Right: M31 radial
profile broken up into three different metallicity bins.  The dip in the metal-rich cluster
distribution between log R = 0.3 and 0.7 ($2<$R$<5$ kpc) appears to be real.
 \label{radialprofile}}
\end{figure}

The left panel of Figure \ref{radialprofile} shows the radial distribution 
of all M31 and MW GCs,
the latter data coming again from the compilation of \citet{harris}. Projection onto the $YZ$ plane
was used for MW clusters, and circular symmetry was assumed for both galaxies.  The outer
M31 clusters ($R> 3$ kpc) follow the $-2.5$ power law seen in the MW halo stars, but the M31 
GC profile is more shallow than that at smaller radii (the exponent changes to   $-1.2$;
overall the M31 GC profile may be fit with a S\'{e}rsic profile with index $n$=2.3).  
It is hard to assess the importance of the difference 
between the two galaxies' profiles in the inner 3 kpc, given that these are the only two galaxies
for which we have such detailed information.
If the M31 clusters
are divided up into three metallicity regimes as shown in the figure, we  see that the metal$-$rich
clusters have a gap in their distribution, with few between 2 and 5 kpc. Several of the metal$-$rich
clusters in
the denser region beyond 5 kpc are possibly not true globulars (e.g., the low mass
clusters B186, B279-D068, B054-G115, and B522), but this represents only a quarter of the clusters in
that region, and thus the gap  between 2 and 5 kpc appears to be real.  The gap does not 
appear for either of the two more metal$-$poor bins at that radius shown in the figure.  

We will continue our comparisons with the MW GC system
when we present the old cluster kinematics in another paper in this
series (A. Romanowsky et al., 2011, in preparation).

\section{Summary and General Comments}

Using the high$-$quality spectra reported here, we
have provided new homogeneous estimates of the metallicities
and ages  for more than 300 globular clusters in M31.  
Again, within a radius 21 kpc we observed 94\%  of the old
clusters.  We note that only 13\% of MW GCs lie
beyond that limit of 21 kpc. The search for outer M31 clusters, presumably metal$-$poor,
is still underway 
\cite[e.g.,][]{huxor}, but it is unlikely that enough clusters will be
discovered to change our basic
comparison of metal$-$rich  and metal$-$poor clusters in M31

We find no evidence for any massive clusters in M31
with intermediate ages, those between 2 and 6 Gyr. 

The metallicities
span the range of those found in the MW, with a few clusters
perhaps extending beyond the most metal$-$rich galactic globulars.
The M31 cluster metallicity distribution is quite different from that of the MW, 
not showing a strong bimodality as does the MW. However, there are hints of 
multi$-$modality. Since it is likely that GCs were deposited in 
M31 in a hierarchical fashion from merging and accretion of a number of 
smaller galaxies, the lack of simple bimodality in the M31 cluster [Fe/H] 
distribution is perhaps not surprising.

Our new data confirm about 320 true GCs.
To that we may add another 50 clusters which we did not observe, but which have been
confirmed as globulars by various sources, but most significantly by 
\cite{barmby} and \cite{huxor}. Thus the total number of known M31 GCs is about 370, compared
with  150 for the MW \citep{harris}. 
We close this paper with some other
comparisons between the two galaxies.

Roughly, in M31 the metal$-$poor and metal$-$rich groups have equal numbers of clusters.  There are 333 clusters
with [Fe/H] measured here or in the literature.  If we divide them at [Fe/H]=$-1$, there are 160 above that with
a combined mass of $1 \times 10^8  M_\sun$ and 173 below it with a mass of $8 \times 10^7 M_\sun$.
If we place all the clusters with unmeasured [Fe/H] into the metal$-$poor group (there are 39 of these,
mostly in the outer halo), 
the result changes only slightly.
There would be 218 clusters in the metal$-$poor group, but the total mass increases only to
 $9 \times 10^7 M_\sun$ since most of the added clusters are low mass.
In the MW, that same dividing point in  [Fe/H] divides the clusters into 1/3 metal$-$rich and 2/3 metal$-$poor,
with the former having a total mass of  $9 \times 10^6  M_\sun$ and 
the latter with $3 \times 10^7  M_\sun$.

The relative total masses of M31 and the MW are a
topic of current interest \citep{reid}, but we can limit the discussion to the
bulge masses, the ratio of which is about a factor of 2 \citep{vdk}.  Therefore, the ratio of
number of GCs of all metallicities to the parent's bulge mass  does appear to be similar in the two galaxies.
However, the ratio of the numbers of metal$-$rich GCs to bulge mass, or equivalently, 
the ratio of metal$-$poor clusters to
the bulge mass is not similar, due implicitly to the different ratio of  metal$-$rich to metal$-$poor clusters between
M31 and the MW.

The specific frequency of GCs is defined as the ratio of
the number of clusters of all metallicity to the galaxy luminosity in units of
$M_{\rm V}=-15$.  Since M31 has about 370 GCs, and
has $M_{\rm V}=-21.2$ \citep{vdb2}, we then find a specific frequency of
1.2.  Isolating the bulge luminosity (taken to be 30\% of the total
light), we derive a bulge specific frequency of 4.  These numbers can be
compared with the values of 1 and 2 for the MW (the latter for the bulge
specific frequency.  Thus, M31 has more clusters per unit luminosity of
old stars.

In M31, 50\% of the known GCs lie within 5.1 kpc  of the Galactic center, remarkably close
to  the  half$-${\it total number} radius of 4.8 kpc for the MW.  (We calculated the 
projected radius of the MW GCs in the $YZ$ plane).
However, the  half$-${\it mass} radius of the M31 GC system is smaller,
3.7 kpc, because the more distant clusters are less massive in the mean than the inner ones.
The half$-${\it mass} radius for the MW GC system  is about 5.5 kpc, slightly 
larger than the half$-${\it total number} radius.

Succeeding papers in this series will concentrate on the abundance ratios
of the clusters, the relation of kinematics and abundances in the cluster
system, the horizontal branch morphologies, and the $M/L$ ratios of the clusters 
as derived from high dispersion
spectroscopy and {\it HST} imaging.

\acknowledgments
We would like to thank Dan Fabricant for leading the effort to
design \& build the Hectospec fiber positioner and spectrograph,
Perry Berlind \&  Mike Calkins for help with
the observations, and Susan Tokarz, John Roll, Maureen Conroy \&  Bill Joye for their many 
contributions to the Hectospec software development project. Important
discussions were had with Pauline Barmby, Paul Hodge, Jay Strader,
Anil Seth and Bob Zinn.
HLM was supported by NSF grant AST-0607518.
Work on this project has also been supported by {\it HST} grant GO10407.

\LongTables

\clearpage
\pagestyle{empty}
\begin{deluxetable}{llllrcrrl}
\tabletypesize{\scriptsize}
\tablecolumns{9}
\tablewidth{0pc}
\tablecaption{Results from the Spectra of Old M31 Clusters \label{main}} 

\tablehead{\colhead{Object} & \colhead{RA}  &\colhead{Dec} &\colhead{$E(B-V)$\tablenotemark{a} }  &\colhead{Velocity}   &\colhead{Log Mass}  &\colhead{[Fe/H]}  &\colhead{Age\tablenotemark{b}} &\colhead{notes\tablenotemark{c}} \\
\colhead{} & \multicolumn{2}{c}{J2000} &\colhead{}    &\colhead{\kms} &\colhead{M$_{\sun}$} &\colhead{} &\colhead{Gyr} &\colhead{} 
}
\startdata
B291-G009&	0:36:04.97&	42:02:09.3&	0.12&	$-$213.5$\pm	14$&	5.5&	$-$1.2$\pm	0.1$&	(14)&		\\
B295-G014&	0:36:46.73&	40:19:42.1&	0.03&	$-$415.3$\pm	26$&	5.3&	$-$1.7$\pm	0.2$&	(14)&		\\
B422&	0:37:38.45&	41:59:59.2&	0.09&	$-$202.4$\pm	18$&	4.9&	$-$1.2$\pm	0.2$&	(14)&		\\
B301-G022&	0:38:21.59&	40:03:37.0&	0.19&	$-$366.6$\pm	7$&	5.5&	$-$0.9$\pm	0.1$&	(14)&		\\
B167D&	0:38:22.48&	41:54:35.0&	0.11&	$-$196.4$\pm	15$&	5.0&	$-$1.5$\pm	0.2$&	(14)&		\\
B302-G023&	0:38:33.5&	41:20:52.2&	0.05&	$-$394.4$\pm	22$&	5.4&	$-$1.5$\pm	0.3$&	(14)&		\\
MCGC7-H14&	0:38:49.39&	42:22:47.0&	(0.13)&	$-$247.6$\pm	25$&	4.9&	$-$0.7$\pm	0.7$&	(14)&		\\
B304-G028&	0:38:56.94&	41:10:28.4&	0.14&	$-$405.3$\pm	15$&	5.4&	$-$1.3$\pm	0.1$&	(14)&		\\
B306-G029&	0:39:08.70&	40:34:21.2&	0.57&	$-$443.9$\pm	9$&	6.1&	$-$1.1$\pm	0.1$&	(14)&		\\
B309-G031&	0:39:24.62&	40:14:29.1&	0.20&	$-$404.3$\pm	16$&	5.3&	$-$1.2$\pm	0.2$&	(14)&		\\
B310-G032&	0:39:25.75&	41:23:33.1&	0.13&	$-$238.1$\pm	14$&	5.3&	$-$1.4$\pm	0.1$&	(14)&		\\
B181D&	0:39:30.85&	41:28:26.4&	0.03&	$-$203.7$\pm	8$&	5.0&	$-$1.0$\pm	0.1$&	(14)&		\\
B311-G033&	0:39:33.72&	40:31:14.7&	0.36&	$-$523.4$\pm	20$&	6.2&	$-$1.9$\pm	0.2$&	(14)&		\\
B312-G035&	0:39:40.17&	40:57:02.4&	0.23&	$-$177.5$\pm	9$&	6.1&	$-$1.2$\pm	0.1$&	(14)&		\\
B313-G036&	0:39:44.60&	40:52:55.2&	0.39&	$-$429.5$\pm	8$&	5.9&	$-$1.0$\pm	0.1$&	(14)&		\\
B001-G039&	0:39:51.02&	40:58:10.6&	0.39&	$-$203.3$\pm	8$&	5.7&	$-$0.7$\pm	0.1$&	11.9&		\\
B317-G041\tablenotemark{j}&	0:39:55.29&	41:47:45.9&	0.12&	$-$178.2$\pm	16$&	5.5&	$-$2.1$\pm	0.2$&	(14)&		\\
B002-G043&	0:40:02.57&	41:11:53.5&	0.11&	$-$338.2$\pm	14$&	5.1&	$-$2.2$\pm	0.2$&	(14)&		\\
B003-G045&	0:40:09.40&	41:11:05.6&	0.12&	$-$377.0$\pm	11$&	5.1&	$-$1.6$\pm	0.2$&	(14)&		\\
B004-G050&	0:40:17.92&	41:22:40.2&	0.18&	$-$369.8$\pm	6$&	5.5&	$-$0.7$\pm	0.1$&	12.2&		\\
B005-G052&	0:40:20.33&	40:43:58.3&	0.16&	$-$291.6$\pm	9$&	5.9&	$-$0.7$\pm	0.1$&	(14)&		\\
B328-G054\tablenotemark{j}&	0:40:24.52&	41:40:23.1&	0.16&	$-$281.5$\pm	13$&	5.2&	$-$1.2$\pm	0.2$&	(14)&	e	\\
B330-G056\tablenotemark{j}&	0:40:25.58&	41:42:53.6&	0.12&	$-$247.3$\pm	17$&	5.1&	$-$1.4$\pm	0.1$&	(14)&		\\
B331-G057\tablenotemark{j}&	0:40:26.10&	41:42:03.9&	0.06&	$-$242.9$\pm	8$&	4.8&	$-$0.8$\pm	0.2$&	5.3&		\\
B006-G058&	0:40:26.48&	41:27:26.7&	0.17&	$-$232.4$\pm	6$&	6.0&	$-$0.5$\pm	0.1$&	(14)&		\\
B244&	0:40:26.49&	41:18:35.5&	0.10&	$-$237.8$\pm	13$&	4.8&	$-$1.4$\pm	0.2$&	(14)&		\\
BH04\tablenotemark{j}&	0:40:27.2&	41:42:23.9&	0.06&	$-$224.6$\pm	22$&	4.2&	$-$0.4$\pm	0.6$&	(14)&		\\
B333\tablenotemark{j}&	0:40:29.58&	41:40:26.7&	0.10&	$-$239.3$\pm	8$&	4.5&	$-$0.8$\pm	0.2$&	5.3&		\\
B008-G060&	0:40:30.28&	41:16:08.7&	0.20&	$-$318.0$\pm	4$&	5.6&	$-$0.8$\pm	0.1$&	13.5&		\\
B009-G061\tablenotemark{j}&	0:40:30.70&	41:36:55.6&	0.10&	$-$294.9$\pm	10$&	5.4&	$-$1.3$\pm	0.1$&	(14)&		\\
B010-G062&	0:40:31.56&	41:14:22.5&	0.19&	$-$164.4$\pm	10$&	5.6&	$-$1.8$\pm	0.2$&	(14)&		\\
B011-G063\tablenotemark{j}&	0:40:31.87&	41:39:16.9&	0.10&	$-$237.8$\pm	13$&	5.4&	$-$1.2$\pm	0.1$&	(14)&		\\
B012-G064&	0:40:32.46&	41:21:44.2&	0.17&	$-$355.7$\pm	11$&	6.2&	$-$1.7$\pm	0.1$&	(14)&		\\
B013-G065&	0:40:38.43&	41:25:23.7&	0.17&	$-$411.5$\pm	6$&	5.3&	$-$0.5$\pm	0.1$&	9.5&		\\
B015-V204&	0:40:45.02&	40:59:56.3&	0.58&	$-$460.0$\pm	7$&	5.6&	$-$0.1$\pm	0.2$&	7.2&		\\
B016-G066&	0:40:45.16&	41:22:09.9&	0.27&	$-$388.4$\pm	5$&	5.3&	$-$0.7$\pm	0.1$&	11.1&		\\
B336-G067&	0:40:47.60&	42:08:43.2&	0.05&	$-$609.0$\pm	27$&	4.9&	$-$2.5$\pm	0.6$&	(14)&		\\
B337-G068&	0:40:48.47&	42:12:11.0&	0.10&	$-$278.5$\pm	14$&	5.4&	$-$1.2$\pm	0.1$&	(14)&		\\
B017-G070&	0:40:48.73&	41:12:07.1&	0.47&	$-$530.4$\pm	6$&	6.2&	$-$0.8$\pm	0.1$&	11.8&		\\
B246&	0:40:52.29&	40:53:55.9&	0.08&	$-$496.8$\pm	13$&	4.7&	$-$0.7$\pm	0.3$&	(14)&		\\
B019-G072&	0:40:52.52&	41:18:53.8&	0.22&	$-$220.4$\pm	6$&	6.3&	$-$0.8$\pm	0.1$&	10.9&		\\
B020-G073&	0:40:55.26&	41:41:25.2&	0.11&	$-$345.4$\pm	5$&	6.2&	$-$0.9$\pm	0.1$&	8.6&		\\
B338-G076&	0:40:58.87&	40:35:47.9&	0.15&	$-$275.8$\pm	12$&	6.5&	$-$1.1$\pm	0.1$&	(14)&		\\
B021-G075&	0:40:58.99&	41:05:39.1&	0.56&	$-$419.1$\pm	7$&	5.6&	$-$0.5$\pm	0.1$&	12.6&		\\
B022-G074&	0:40:59.08&	41:24:42.0&	0.14&	$-$454.4$\pm	11$&	5.2&	$-$1.8$\pm	0.2$&	(14)&		\\
B339-G077&	0:41:00.71&	39:55:54.2&	(0.13)&	$-$227.1$\pm	13$&	5.4&	$-$0.5$\pm	0.2$&	(14)&		\\
B023-G078&	0:41:01.19&	41:13:45.7&	0.42&	$-$443.9$\pm	7$&	6.9&	$-$0.7$\pm	0.1$&	11.8&		\\
B248&	0:41:07.94&	40:53:01.0&	0.30&	$-$570.3$\pm	10$&	5.2&	$-$1.4$\pm	0.2$&	(14)&		\\
B341-G081&	0:41:09.15&	40:35:52.8&	0.20&	$-$364.9$\pm	8$&	5.7&	$-$0.9$\pm	0.1$&	12.4&		\\
B024-G082&	0:41:11.86&	41:45:49.1&	0.19&	$-$215.0$\pm	5$&	5.5&	$-$0.6$\pm	0.1$&	(14)&		\\
B025-G084&	0:41:12.55&	41:00:28.3&	0.35&	$-$204.9$\pm	15$&	5.7&	$-$1.4$\pm	0.1$&	(14)&		\\
B249&	0:41:12.58&	41:01:12.7&	0.37&	$-$536.7$\pm	9$&	5.2&	0.3$\pm	0.2$&	8.5&		\\
B027-G087&	0:41:14.54&	40:55:50.9&	0.19&	$-$302.9$\pm	16$&	6.0&	$-$1.3$\pm	0.1$&	(14)&		\\
B026-G086&	0:41:14.55&	41:24:40.1&	0.28&	$-$250.8$\pm	5$&	5.3&	$-$0.4$\pm	0.1$&	(14)&		\\
B028-G088&	0:41:16.50&	40:59:03.2&	0.20&	$-$414.2$\pm	19$&	5.5&	$-$2.5$\pm	0.2$&	(14)&		\\
B020D-G089&	0:41:17.23&	41:08:09.1&	0.49&	$-$558.0$\pm	17$&	5.6&	$-$2.1$\pm	0.2$&	(14)&		\\
B029-G090&	0:41:17.82&	41:00:23.0&	0.27&	$-$500.0$\pm	7$&	5.7&	$-$0.3$\pm	0.1$&	13.4&		\\
B030-G091&	0:41:18.74&	40:57:15.6&	0.65&	$-$390.9$\pm	6$&	5.9&	$-$0.3$\pm	0.1$&	(14)&		\\
B031-G092&	0:41:20.93&	40:59:04.2&	0.31&	$-$310.8$\pm	8$&	5.3&	$-$0.5$\pm	0.1$&	11.5&		\\
B032-G093&	0:41:21.51&	41:17:30.2&	0.51&	$-$526.3$\pm	7$&	5.6&	$-$0.7$\pm	0.1$&	12.3&		\\
B033-G095&	0:41:26.40&	41:00:14.0&	0.30&	$-$466.5$\pm	16$&	5.3&	$-$1.6$\pm	0.3$&	(14)&		\\
B034-G096&	0:41:28.12&	40:53:49.6&	0.16&	$-$556.5$\pm	10$&	6.0&	$-$0.6$\pm	0.1$&	(14)&		\\
B457-G097&	0:41:29.23&	42:18:37.1&	0.13&	$-$330.0$\pm	18$&	5.4&	$-$1.1$\pm	0.2$&	(14)&		\\
B035&	0:41:32.58&	41:38:32.7&	0.21&	$-$45.7$\pm	5$&	5.3&	$-$0.8$\pm	0.1$&	11.7&		\\
B036&	0:41:32.83&	41:26:05.1&	0.25&	$-$506.2$\pm	6$&	5.4&	$-$0.8$\pm	0.1$&	(14)&		\\
B037-V327&	0:41:34.98&	41:14:55.1&	1.61&	$-$322.8$\pm	11$&	7.2&	$-$0.8$\pm	0.1$&	(14)&		\\
B038-G098&	0:41:35.95&	41:19:14.8&	0.40&	$-$183.4$\pm	9$&	5.9&	$-$1.7$\pm	0.1$&	(14)&		\\
B039-G101&	0:41:37.87&	41:20:50.1&	0.60&	$-$246.6$\pm	6$&	6.3&	$-$0.8$\pm	0.1$&	9.0&		\\
B041-G103\tablenotemark{i}&	0:41:40.81&	41:14:45.4&	(0.13)&	$-$398.2$\pm	2$&	4.7&		&	(14)&		\\
B042-G104&	0:41:41.69&	41:07:26.2&	0.88&	$-$293.8$\pm	12$&	6.6&	$-$0.9$\pm	0.1$&	6.7&		\\
B044-G107&	0:41:42.91&	41:20:06.2&	0.47&	$-$267.6$\pm	6$&	5.8&	$-$0.8$\pm	0.1$&	13.3&		\\
B343-G105&	0:41:43.10&	40:12:22.4&	0.15&	$-$356.0$\pm	14$&	5.7&	$-$1.6$\pm	0.2$&	(14)&		\\
B045-G108&	0:41:43.11&	41:34:20.3&	0.18&	$-$419.4$\pm	6$&	5.9&	$-$0.9$\pm	0.1$&	(14)&		\\
B046-G109&	0:41:44.60&	41:46:27.7&	0.18&	$-$351.3$\pm	7$&	5.1&	$-$1.1$\pm	0.1$&	(14)&		\\
B048-G110&	0:41:45.53&	41:13:30.6&	0.36&	$-$234.6$\pm	10$&	5.8&	$-$0.8$\pm	0.1$&	(14)&		\\
B047-G111&	0:41:45.56&	41:42:04.1&	0.16&	$-$294.8$\pm	9$&	5.2&	$-$1.6$\pm	0.2$&	(14)&		\\
B050-G113&	0:41:46.28&	41:32:18.7&	0.25&	$-$109.5$\pm	6$&	5.6&	$-$0.8$\pm	0.1$&	13.5&		\\
B051-G114&	0:41:46.70&	41:25:19.1&	0.38&	$-$261.2$\pm	6$&	6.0&	$-$0.8$\pm	0.1$&	(14)&		\\
B054-G115&	0:41:47.68&	41:00:55.3&	0.17&	$-$397.2$\pm	8$&	5.0&	$-$0.2$\pm	0.1$&	10.6&		\\
B055-G116&	0:41:50.39&	41:12:12.4&	0.54&	$-$304.9$\pm	8$&	6.0&	$-$0.1$\pm	0.1$&	(14)&		\\
B254&	0:41:50.5&	41:16:25.9&	0.29&	$-$421.0$\pm	12$&	4.9&	$-$0.9$\pm	0.2$&	4.8&		\\
B522&	0:41:50.95&	40:52:48.2&	0.11&	$-$390.2$\pm	23$&	4.7&	$-$0.4$\pm	0.6$&	9.1&	e	\\
B056-G117&	0:41:51.16&	40:57:40.2&	0.24&	$-$363.7$\pm	7$&	5.4&	$-$0.0$\pm	0.1$&	(14)&		\\
B057-G118&	0:41:52.82&	40:52:05.1&	0.10&	$-$411.1$\pm	21$&	5.1&	$-$2.1$\pm	0.3$&	(14)&		\\
B058-G119&	0:41:53.00&	40:47:09.7&	0.15&	$-$220.2$\pm	11$&	6.2&	$-$1.1$\pm	0.1$&	(14)&		\\
B059-G120&	0:41:54.11&	41:11:00.7&	0.49&	$-$262.1$\pm	8$&	5.7&	$-$1.0$\pm	0.1$&	(14)&		\\
B060-G121&	0:41:57.01&	41:05:14.5&	0.13&	$-$533.2$\pm	16$&	5.5&	$-$1.8$\pm	0.1$&	(14)&		\\
B061-G122&	0:42:00.14&	41:29:35.7&	0.49&	$-$275.2$\pm	5$&	6.0&	$-$0.7$\pm	0.1$&	(14)&		\\
B063-G124&	0:42:00.88&	41:29:09.5&	0.49&	$-$301.6$\pm	8$&	6.3&	$-$0.8$\pm	0.1$&	13.8&		\\
B065-G126&	0:42:01.93&	40:40:13.1&	0.17&	$-$417.3$\pm	12$&	5.5&	$-$0.9$\pm	0.2$&	(14)&		\\
B064-G125&	0:42:01.93&	41:11:07.5&	0.17&	$-$302.4$\pm	10$&	5.7&	$-$1.3$\pm	0.1$&	(14)&		\\
B344-G127&	0:42:02.97&	41:52:02.2&	0.13&	$-$240.6$\pm	6$&	5.8&	$-$1.0$\pm	0.1$&	(14)&		\\
B067-G129&	0:42:03.19&	41:04:23.7&	0.08&	$-$356.8$\pm	15$&	5.2&	$-$1.4$\pm	0.1$&	(14)&		\\
B068-G130&	0:42:03.21&	40:58:50.2&	0.45&	$-$319.2$\pm	6$&	6.1&	$-$0.2$\pm	0.1$&	(14)&		\\
B257-V219&	0:42:03.28&	40:58:13.9&	0.66&	$-$476.4$\pm	6$&	5.7&	$-$0.2$\pm	0.1$&	(14)&		\\
B461-G131&	0:42:04.24&	42:03:26.6&	0.10&	$-$296.5$\pm	8$&	5.1&	$-$0.6$\pm	0.1$&	13.7&		\\
B041D&	0:42:04.72&	41:16:47.3&	0.64&	$-$208.3$\pm	14$&	5.5&	$-$1.8$\pm	0.4$&	(14)&		\\
B070-G133\tablenotemark{d}&	0:42:06.91&	41:07:56.3&	0.20&	$-$275.8$\pm	22$&	5.6&	$-$1.9$\pm	0.2$&	(14)&		\\
B071&	0:42:07.13&	41:12:12.0&	0.24&	$-$550.7$\pm	8$&	5.1&	$-$0.2$\pm	0.2$&	6.9&		\\
B073-G134&	0:42:07.33&	40:59:21.3&	0.13&	$-$504.0$\pm	10$&	5.8&	$-$0.6$\pm	0.1$&	12.6&		\\
B072&	0:42:07.44&	41:22:47.6&	0.78&	$-$85.7$\pm	7$&	5.8&	$-$0.7$\pm	0.1$&	12.0&		\\
B074-G135&	0:42:08.04&	41:43:21.6&	0.16&	$-$439.9$\pm	9$&	5.5&	$-$1.5$\pm	0.1$&	(14)&		\\
B075-G136&	0:42:08.83&	41:20:21.3&	0.32&	$-$168.6$\pm	11$&	5.4&	$-$1.3$\pm	0.2$&	(14)&		\\
MITA140&	0:42:09.51&	41:17:45.6&	0.86&	$-$319.4$\pm	12$&	6.3&	$-$0.2$\pm	0.1$&	(14)&		\\
B045D&	0:42:09.87&	41:21:14.5&	0.41&	$-$305.9$\pm	13$&	4.9&	$-$0.3$\pm	0.3$&	(14)&		\\
B076-G138&	0:42:10.24&	41:05:22.0&	0.17&	$-$537.0$\pm	14$&	5.5&	$-$1.3$\pm	0.1$&	(14)&		\\
B077-G139&	0:42:11.14&	41:07:33.9&	0.46&	$-$562.8$\pm	7$&	5.7&	$-$0.8$\pm	0.1$&	(14)&	e	\\
B078-G140&	0:42:12.17&	41:17:58.9&	(0.43)&	$-$259.7$\pm	16$&	5.5&	$-$1.0$\pm	0.2$&	(14)&		\\
B080-G141&	0:42:12.40&	41:19:00.6&	0.72&	$-$255.0$\pm	21$&	5.8&	$-$1.4$\pm	0.2$&	(14)&		\\
B345-G143&	0:42:14.12&	40:17:36.5&	0.19&	$-$368.4$\pm	12$&	5.6&	$-$1.5$\pm	0.1$&	(14)&		\\
B462&	0:42:14.72&	42:01:36.7&	0.13&	$-$214.3$\pm	23$&	4.9&	$-$2.8$\pm	0.3$&	(14)&		\\
B082-G144&	0:42:15.84&	41:01:14.4&	0.94&	$-$370.5$\pm	8$&	6.7&	$-$0.7$\pm	0.1$&	10.3&		\\
B083-G146&	0:42:16.44&	41:45:20.7&	0.13&	$-$337.3$\pm	6$&	5.3&	$-$1.1$\pm	0.1$&	(14)&		\\
B084&	0:42:17.45&	41:18:55.7&	0.62&	$-$325.8$\pm	10$&	5.6&	$-$1.1$\pm	0.2$&	(14)&		\\
B085-G147&	0:42:18.24&	40:39:57.2&	0.16&	$-$425.4$\pm	12$&	5.5&	$-$1.7$\pm	0.2$&	(14)&		\\
B086-G148&	0:42:18.65&	41:14:02.1&	0.13&	$-$189.0$\pm	19$&	6.1&	$-$1.8$\pm	0.2$&	(14)&	e	\\
B087&	0:42:19.81&	41:38:16.2&	0.24&	$-$369.2$\pm	9$&	4.9&	$-$1.0$\pm	0.2$&	(14)&	near star	\\
B088-G150&	0:42:21.07&	41:32:14.2&	0.53&	$-$489.9$\pm	12$&	6.5&	$-$1.8$\pm	0.1$&	(14)&		\\
B090&	0:42:21.08&	41:02:57.5&	0.15&	$-$399.9$\pm	7$&	4.8&	$-$0.8$\pm	0.2$&	(14)&	e	\\
B092-G152&	0:42:22.38&	41:08:08.7&	0.11&	$-$433.7$\pm	12$&	5.4&	$-$1.0$\pm	0.1$&	(14)&		\\
B347-G154&	0:42:22.89&	41:54:27.5&	0.02&	$-$266.4$\pm	26$&	5.4&	$-$2.0$\pm	0.2$&	(14)&		\\
B348-G153&	0:42:22.92&	41:52:28.4&	0.16&	$-$189.5$\pm	7$&	5.5&	$-$0.8$\pm	0.1$&	(14)&		\\
B093-G155&	0:42:23.17&	41:21:43.5&	0.39&	$-$456.7$\pm	22$&	5.7&	$-$1.2$\pm	0.1$&	(14)&		\\
B094-G156&	0:42:25.06&	40:57:17.7&	0.22&	$-$564.1$\pm	6$&	6.0&	$-$0.4$\pm	0.1$&	(14)&		\\
B095-G157&	0:42:25.80&	41:05:36.3&	0.46&	$-$111.2$\pm	10$&	6.1&	$-$1.4$\pm	0.1$&	(14)&		\\
B096-G158&	0:42:26.1&	41:19:14.8&	0.63&	$-$313.4$\pm	8$&	6.1&	$-$0.3$\pm	0.1$&	(14)&		\\
B098&	0:42:27.40&	40:59:36.1&	0.19&	$-$308.6$\pm	9$&	5.7&	$-$0.8$\pm	0.1$&	8.7&		\\
B097-G159&	0:42:27.48&	41:25:32.1&	0.34&	$-$272.1$\pm	8$&	5.7&	$-$1.0$\pm	0.1$&	(14)&		\\
B099-G161&	0:42:27.59&	41:10:02.7&	0.16&	$-$108.1$\pm	10$&	5.5&	$-$1.3$\pm	0.1$&	(14)&		\\
B515\tablenotemark{i}&	0:42:28.05&	41:33:24.5&	(0.13)&	$-$276.7$\pm	2$&	4.7&		&	(14)&		\\
B056D&	0:42:28.36&	41:34:27.2&	(0.13)&	$-$188.0$\pm	12$&	4.8&		&	&	r	\\
B350-G162&	0:42:28.44&	40:24:51.1&	0.12&	$-$423.9$\pm	12$&	5.5&	$-$1.4$\pm	0.1$&	(14)&		\\
B100-G163&	0:42:28.96&	40:49:56.0&	0.15&	$-$428.4$\pm	8$&	5.1&	$-$0.9$\pm	0.1$&	(14)&		\\
B101-G164&	0:42:29.04&	41:08:15.6&	0.13&	$-$352.0$\pm	9$&	5.4&	$-$1.0$\pm	0.1$&	(14)&		\\
B103-G165&	0:42:29.75&	41:17:57.5&	0.34&	$-$366.5$\pm	7$&	6.4&	$-$0.5$\pm	0.1$&	13.3&		\\
B104-NB5&	0:42:29.94&	41:17:25.7&	0.20&	45.6$\pm	22$&	5.2&	$-$1.4$\pm	0.2$&	(14)&	e	\\
B105-G166&	0:42:30.75&	41:30:27.3&	0.16&	$-$240.2$\pm	7$&	5.3&	$-$1.0$\pm	0.1$&	(14)&	two objects	\\
B106-G168&	0:42:31.04&	41:12:18.3&	0.22&	$-$62.6$\pm	7$&	5.7&	$-$0.6$\pm	0.1$&	13.2&		\\
B108-G167&	0:42:31.19&	41:08:51.3&	0.22&	$-$548.0$\pm	11$&	5.4&	0.0$\pm	0.2$&	10.4&		\\
B107-G169&	0:42:31.27&	41:19:38.9&	0.22&	$-$332.4$\pm	7$&	6.0&	$-$1.0$\pm	0.1$&	(14)&		\\
B109-G170&	0:42:32.16&	41:10:27.9&	0.20&	$-$612.1$\pm	8$&	5.7&	$-$0.2$\pm	0.1$&	9.0&		\\
B110-G172&	0:42:33.10&	41:03:28.4&	0.12&	$-$239.4$\pm	10$&	6.1&	$-$0.7$\pm	0.1$&	8.6&		\\
NB16&	0:42:33.12&	41:20:16.8&	0.68&	$-$95.1$\pm	36$&	5.5&	$-$1.4$\pm	0.5$&	(14)&		\\
B111-G173&	0:42:33.17&	41:00:26.5&	0.15&	$-$402.5$\pm	15$&	5.5&	$-$1.3$\pm	0.1$&	(14)&		\\
B260&	0:42:33.19&	41:31:24.8&	1.00&	$-$189.8$\pm	17$&	5.8&	$-$0.6$\pm	0.3$&	(14)&	e	\\
B112-G174&	0:42:33.26&	41:17:42.4&	(0.13)&	$-$272.6$\pm	11$&	5.5&	0.3$\pm	0.1$&	(14)&		\\
B114-G175&	0:42:34.30&	41:12:44.9&	0.19&	$-$244.5$\pm	30$&	5.4&	$-$2.2$\pm	0.4$&	(14)&		\\
B117-G176&	0:42:34.38&	40:57:09.3&	0.08&	$-$527.8$\pm	20$&	5.3&	$-$1.7$\pm	0.2$&	(14)&		\\
B115-G177&	0:42:34.41&	41:14:02.0&	0.18&	$-$593.4$\pm	11$&	5.8&	0.1$\pm	0.1$&	(14)&	e	\\
B116-G178&	0:42:34.54&	41:32:51.4&	0.72&	$-$339.3$\pm	7$&	6.2&	$-$0.6$\pm	0.1$&	(14)&		\\
B064D-NB6&	0:42:35.54&	41:14:34.3&	0.14&	21.5$\pm	15$&	5.6&	$-$1.1$\pm	0.2$&	(14)&		\\
B119-NB14&	0:42:36.11&	41:17:35.4&	0.21&	$-$370.4$\pm	15$&	5.3&	$-$0.8$\pm	0.3$&	13.6&	e	\\
NB21-AU5\tablenotemark{h}&	0:42:37.98&	41:15:58.9&	0.02&	$-$632.4$\pm	30$&	4.7&	$-$1.1$\pm	0.3$&	(14)&	e	\\
B351-G179&	0:42:37.98&	42:11:30.7&	0.13&	$-$325.1$\pm	22$&	5.1&	$-$1.5$\pm	0.6$&	(14)&		\\
B352-G180&	0:42:38.19&	42:02:13.1&	0.17&	$-$291.4$\pm	11$&	5.6&	$-$1.5$\pm	0.2$&	(14)&		\\
B068D&	0:42:39.9&	41:20:39.9&	0.08&	$-$202.2$\pm	12$&	4.6&	$-$0.5$\pm	0.3$&	8.7&		\\
B122-G181&	0:42:40.11&	41:33:46.8&	0.79&	$-$436.2$\pm	11$&	5.9&	$-$1.2$\pm	0.1$&	(14)&		\\
B123-G182&	0:42:40.66&	41:10:33.4&	0.20&	$-$368.2$\pm	12$&	5.3&	$-$0.7$\pm	0.2$&	13.0&		\\
B124-NB10&	0:42:41.44&	41:15:23.7&	0.17&	$-$25.5$\pm	10$&	6.3&	$-$0.5$\pm	0.1$&	10.6&		\\
B125-G183&	0:42:42.27&	41:05:31.0&	0.11&	$-$656.8$\pm	12$&	5.5&	$-$1.5$\pm	0.1$&	(14)&		\\
DAO55&	0:42:42.5&	40:29:27.0&	0.14&	$-$447.9$\pm	28$&	4.7&	0.3$\pm	0.5$&	(14)&		\\
B126-G184&	0:42:43.70&	41:12:42.8&	0.13&	$-$188.4$\pm	16$&	5.3&	$-$1.5$\pm	0.2$&	(14)&		\\
B127-G185&	0:42:44.50&	41:14:41.5&	0.20&	$-$528.5$\pm	7$&	6.5&	$-$0.7$\pm	0.1$&	9.5&	e	\\
SK054A&	0:42:45.08&	41:08:15.1&	(0.13)&	$-$627.9$\pm	11$&	4.8&		&	&	r	\\
B072D&	0:42:45.79&	41:27:27.0&	0.16&	$-$262.5$\pm	12$&	4.6&	$-$0.9$\pm	0.2$&	8.6&		\\
BH16\tablenotemark{i}&	0:42:46.10&	41:17:36.2&	(0.13)&	$-$99.9$\pm	1$&	4.7&		&	(14)&		\\
NB18&	0:42:46.34&	41:18:32.4&	0.00&	$-$210.2$\pm	39$&	4.5&		&	(14)&	w	\\
B354-G186&	0:42:47.64&	42:00:24.7&	0.15&	$-$172.6$\pm	17$&	5.1&	$-$1.6$\pm	0.4$&	(14)&		\\
B128-G187&	0:42:47.81&	41:11:13.8&	(0.24)&	$-$378.8$\pm	10$&	5.5&	$-$0.6$\pm	0.1$&	13.7&	e	\\
B129&	0:42:48.35&	41:25:06.6&	1.24&	$-$42.9$\pm	11$&	6.7&	$-$0.8$\pm	0.1$&	8.2&		\\
B130-G188&	0:42:48.86&	41:29:52.7&	0.42&	$-$22.6$\pm	10$&	5.8&	$-$1.2$\pm	0.1$&	(14)&		\\
B262\tablenotemark{g}&	0:42:50.05&	41:19:28.1&	0.06&	$-$378.4$\pm	21$&	5.0&	$-$1.3$\pm	0.3$&	(14)&		\\
BH18&	0:42:50.73&	41:10:33.4&	0.21&	$-$517.7$\pm	14$&	5.0&	0.1$\pm	0.3$&	10.1&		\\
B131-G189&	0:42:50.81&	41:17:07.3&	0.18&	$-$467.5$\pm	10$&	6.1&	$-$0.7$\pm	0.1$&	10.0&		\\
B132-NB15\tablenotemark{f}&	0:42:51.44&	41:15:40.7&	0.30&	36.2$\pm	18$&	5.3&	$-$0.5$\pm	0.3$&	(14)&	e	\\
B134-G190&	0:42:51.65&	41:14:03.6&	0.13&	$-$374.7$\pm	8$&	5.5&	$-$0.9$\pm	0.1$&	(14)&		\\
B078D&	0:42:51.91&	41:22:05.2&	0.49&	$-$36.4$\pm	14$&	4.9&	$-$0.4$\pm	0.3$&	(14)&		\\
B135-G192&	0:42:51.98&	41:31:08.3&	0.28&	$-$375.0$\pm	13$&	6.0&	$-$1.8$\pm	0.1$&	(14)&		\\
B264-NB19&	0:42:53.19&	41:16:14.4&	0.21&	$-$49.0$\pm	39$&	5.2&	$-$2.4$\pm	0.9$&	(14)&		\\
B136-G194&	0:42:53.64&	41:19:34.4&	0.12&	$-$627.6$\pm	17$&	5.4&	$-$1.2$\pm	0.2$&	(14)&		\\
B137-G195&	0:42:54.0&	41:32:14.4&	0.52&	$-$220.5$\pm	10$&	5.6&	$-$1.5$\pm	0.1$&	(14)&		\\
B138&	0:42:55.62&	41:18:35.1&	0.22&	$-$347.2$\pm	10$&	5.5&	$-$0.0$\pm	0.2$&	7.4&		\\
AU010&	0:42:58.13&	41:16:52.7&	0.22&	$-$307.3$\pm	12$&	5.3&	$-$0.5$\pm	0.3$&	6.0&		\\
B140-G196&	0:42:58.75&	41:08:52.7&	0.19&	$-$474.7$\pm	11$&	5.1&	$-$0.1$\pm	0.2$&	7.0&		\\
B087D&	0:42:58.92&	41:09:08.8&	0.14&	$-$661.4$\pm	9$&	5.2&	$-$1.2$\pm	0.1$&	(14)&		\\
B141-G197&	0:42:59.29&	41:32:47.5&	0.32&	$-$173.2$\pm	11$&	5.7&	$-$1.4$\pm	0.1$&	(14)&		\\
B088D\tablenotemark{i}&	0:42:59.38&	41:04:17.5&	(0.13)&	$-$346.6$\pm	1$&	&		&	&		\\
B143-G198&	0:42:59.66&	41:19:19.3&	0.34&	$-$144.5$\pm	7$&	6.0&	$-$0.1$\pm	0.1$&	12.4&		\\
B144&	0:42:59.87&	41:16:05.7&	0.24&	$-$6.4$\pm	11$&	5.6&	0.1$\pm	0.2$&	11.7&		\\
B090D&	0:43:01.23&	41:16:10.4&	(0.13)&	15.2$\pm	12$&	5.2&	$-$0.3$\pm	0.2$&	12.4&		\\
B091D-D058&	0:43:01.44&	41:30:17.5&	0.24&	$-$122.8$\pm	6$&	6.1&	$-$0.7$\pm	0.1$&	12.8&		\\
B145&	0:43:01.59&	41:12:26.9&	0.14&	$-$303.0$\pm	17$&	4.8&	$-$1.2$\pm	0.3$&	(14)&		\\
B092D&	0:43:01.70&	41:13:08.9&	0.36&	$-$321.0$\pm	11$&	4.9&	0.4$\pm	0.5$&	(14)&		\\
B265&	0:43:01.92&	40:53:01.8&	0.16&	$-$490.9$\pm	9$&	4.8&	$-$0.8$\pm	0.2$&	13.7&		\\
B146&	0:43:02.94&	41:15:22.6&	0.18&	$-$14.6$\pm	11$&	5.4&	$-$1.0$\pm	0.2$&	(14)&		\\
B147-G199&	0:43:03.30&	41:21:21.5&	(0.13)&	$-$85.3$\pm	11$&	5.9&	$-$0.1$\pm	0.1$&	(14)&		\\
B266&	0:43:03.52&	41:40:31.2&	0.52&	$-$160.6$\pm	8$&	5.3&	$-$1.0$\pm	0.2$&	(14)&		\\
B148-G200&	0:43:03.87&	41:18:04.8&	0.28&	$-$333.6$\pm	8$&	6.0&	$-$1.1$\pm	0.1$&	(14)&		\\
B149-G201&	0:43:05.48&	41:34:27.3&	0.37&	$-$58.0$\pm	8$&	5.7&	$-$1.3$\pm	0.1$&	(14)&		\\
B467-G202&	0:43:06.45&	42:01:49.1&	0.09&	$-$282.4$\pm	17$&	5.1&	$-$1.4$\pm	0.2$&	(14)&		\\
B268&	0:43:07.19&	41:11:47.8&	0.29&	$-$233.0$\pm	14$&	5.0&	$-$0.2$\pm	0.3$&	7.6&		\\
B269&	0:43:07.38&	41:27:32.9&	0.51&	$-$320.3$\pm	10$&	5.1&	$-$0.8$\pm	0.2$&	9.7&	e	\\
B150-G203&	0:43:07.52&	41:20:19.6&	0.19&	$-$138.4$\pm	7$&	5.6&	$-$0.3$\pm	0.1$&	12.7&		\\
PHF6-1&	0:43:08.02&	41:18:18.3&	(0.13)&	$-$116.7$\pm	11$&	5.1&	$-$0.2$\pm	0.2$&	(14)&		\\
B151-G205&	0:43:09.56&	41:21:32.1&	0.53&	$-$340.4$\pm	9$&	6.8&	$-$0.6$\pm	0.1$&	12.8&		\\
B152-G207&	0:43:10.02&	41:18:16.1&	0.18&	$-$136.1$\pm	12$&	5.8&	$-$0.7$\pm	0.1$&	11.2&		\\
B356-G206&	0:43:10.36&	41:50:31.3&	0.27&	$-$186.2$\pm	16$&	5.6&	$-$1.4$\pm	0.1$&	(14)&		\\
B153&	0:43:10.63&	41:14:51.4&	0.21&	$-$248.9$\pm	10$&	5.8&	$-$0.3$\pm	0.1$&	11.8&		\\
B154-G208&	0:43:12.46&	41:16:04.9&	0.17&	$-$209.0$\pm	11$&	5.5&	$-$0.2$\pm	0.2$&	11.7&		\\
B155-G210&	0:43:13.39&	41:03:28.3&	0.19&	$-$416.9$\pm	7$&	5.1&	$-$0.5$\pm	0.1$&	(14)&		\\
B156-G211&	0:43:13.73&	41:01:17.9&	0.09&	$-$373.9$\pm	8$&	5.4&	$-$1.2$\pm	0.1$&	(14)&		\\
B157-G212&	0:43:14.00&	41:11:19.7&	(0.09)&	4.0$\pm	40$&	5.1&	$-$2.6$\pm	0.3$&	(14)&		\\
B158-G213&	0:43:14.41&	41:07:21.2&	0.16&	$-$177.7$\pm	10$&	6.3&	$-$0.8$\pm	0.1$&	9.1&	e	\\
B159&	0:43:14.65&	41:25:13.5&	0.54&	$-$513.3$\pm	14$&	5.7&	$-$1.2$\pm	0.2$&	(14)&		\\
B160-G214&	0:43:14.93&	41:01:35.6&	0.09&	$-$340.1$\pm	21$&	4.9&	$-$2.8$\pm	0.4$&	(14)&		\\
B161-G215&	0:43:15.43&	41:11:25.0&	0.18&	$-$442.7$\pm	12$&	5.7&	$-$1.1$\pm	0.1$&	(14)&		\\
SK063A&	0:43:16.09&	41:27:57.0&	(0.13)&	$-$266.5$\pm	10$&	4.8&		&	&	r	\\
B162-G216&	0:43:16.41&	41:24:04.5&	0.21&	$-$139.4$\pm	8$&	5.2&	$-$0.5$\pm	0.1$&	(14)&	e	\\
B163-G217&	0:43:17.64&	41:27:45.0&	0.21&	$-$161.8$\pm	6$&	6.3&	$-$0.1$\pm	0.1$&	13.5&		\\
B164-V253&	0:43:18.14&	41:12:29.3&	0.23&	50.9$\pm	15$&	5.2&	$-$0.3$\pm	0.2$&	(14)&		\\
B165-G218&	0:43:18.22&	41:10:54.7&	0.15&	$-$68.0$\pm	14$&	5.6&	$-$2.0$\pm	0.2$&	(14)&		\\
B167&	0:43:21.13&	41:14:08.3&	0.18&	$-$199.8$\pm	8$&	5.3&	$-$0.4$\pm	0.1$&	11.7&		\\
B168&	0:43:22.52&	41:44:05.6&	0.76&	$-$117.1$\pm	7$&	5.6&	$-$0.6$\pm	0.1$&	12.4&		\\
B169&	0:43:23.00&	41:15:25.4&	(0.44)&	$-$127.5$\pm	12$&	5.6&	0.3$\pm	0.1$&	(14)&		\\
B170-G221&	0:43:23.47&	40:50:41.8&	0.21&	$-$289.3$\pm	8$&	5.3&	$-$0.7$\pm	0.1$&	(14)&		\\
B272-V294&	0:43:25.52&	41:37:11.7&	0.46&	$-$116.9$\pm	9$&	5.3&	$-$0.9$\pm	0.2$&	(14)&		\\
B171-G222&	0:43:25.61&	41:15:37.2&	0.19&	$-$264.1$\pm	10$&	6.1&	$-$0.3$\pm	0.1$&	12.9&		\\
B172-G223&	0:43:26.00&	41:21:31.6&	0.17&	$-$271.9$\pm	10$&	5.5&	$-$0.6$\pm	0.1$&	10.7&	e	\\
B173-G224&	0:43:28.76&	41:22:37.0&	0.11&	$-$291.5$\pm	8$&	5.2&	$-$0.8$\pm	0.1$&	10.3&		\\
B174-G226&	0:43:30.31&	41:38:56.2&	0.28&	$-$485.7$\pm	12$&	6.2&	$-$1.0$\pm	0.1$&	(14)&		\\
B176-G227&	0:43:30.45&	40:49:11.1&	0.13&	$-$538.8$\pm	13$&	5.4&	$-$1.6$\pm	0.2$&	(14)&		\\
B177-G228&	0:43:30.49&	41:05:42.4&	0.18&	$-$403.0$\pm	9$&	4.9&	$-$1.2$\pm	0.1$&	(14)&		\\
B178-G229&	0:43:30.79&	41:21:16.4&	0.10&	$-$151.4$\pm	14$&	6.1&	$-$1.2$\pm	0.1$&	(14)&		\\
B179-G230&	0:43:31.10&	41:18:14.7&	0.10&	$-$144.6$\pm	11$&	6.0&	$-$1.0$\pm	0.1$&	(14)&		\\
B180-G231&	0:43:31.72&	41:07:46.4&	0.19&	$-$195.2$\pm	5$&	5.7&	$-$0.9$\pm	0.1$&	11.1&		\\
B181-G232&	0:43:32.46&	41:29:07.4&	0.13&	$-$256.5$\pm	10$&	5.4&	$-$0.5$\pm	0.1$&	12.3&		\\
B182-G233&	0:43:36.66&	41:08:12.2&	0.33&	$-$350.6$\pm	8$&	6.2&	$-$1.0$\pm	0.1$&	(14)&		\\
B183-G234&	0:43:36.94&	41:02:02.4&	0.19&	$-$183.2$\pm	6$&	5.8&	$-$0.5$\pm	0.1$&	(14)&		\\
B185-G235&	0:43:37.28&	41:14:43.6&	0.21&	$-$152.9$\pm	7$&	6.0&	$-$0.6$\pm	0.1$&	14.0&		\\
B184-G236&	0:43:37.52&	41:36:34.5&	0.27&	$-$154.0$\pm	7$&	5.3&	0.1$\pm	0.1$&	(14)&		\\
B186&	0:43:38.23&	41:36:24.1&	0.25&	$-$119.2$\pm	10$&	4.7&	0.2$\pm	0.3$&	(14)&	e, diffuse	\\
B187-G237&	0:43:38.64&	41:29:47.1&	0.36&	$-$65.5$\pm	18$&	5.5&	$-$1.6$\pm	0.3$&	(14)&		\\
B188-G239&	0:43:41.51&	41:24:25.6&	0.09&	$-$211.6$\pm	14$&	5.3&	$-$1.7$\pm	0.2$&	(14)&		\\
B189-G240&	0:43:42.42&	41:35:23.3&	0.24&	$-$136.9$\pm	12$&	5.5&	0.4$\pm	0.2$&	(14)&		\\
B190-G241&	0:43:43.39&	41:34:06.0&	0.25&	$-$91.8$\pm	11$&	5.5&	$-$1.2$\pm	0.1$&	(14)&		\\
B194-G243&	0:43:45.18&	41:06:08.7&	0.19&	$-$400.0$\pm	12$&	5.3&	$-$1.4$\pm	0.1$&	(14)&	e	\\
B193-G244&	0:43:45.52&	41:36:57.6&	0.23&	$-$60.8$\pm	6$&	6.2&	$-$0.1$\pm	0.1$&	12.9&		\\
SK072A&	0:43:46.69&	41:22:28.2&	(0.13)&	$-$137.2$\pm	5$&	5.1&		&	&	r	\\
B103D-G245&	0:43:47.54&	41:27:08.0&	0.18&	$-$170.9$\pm	8$&	5.1&	$-$0.8$\pm	0.1$&	10.9&		\\
B472-D064&	0:43:48.42&	41:26:53.2&	0.12&	$-$117.5$\pm	13$&	6.1&	$-$1.0$\pm	0.1$&	(14)&		\\
SK073A&	0:43:48.52&	41:07:48.4&	(0.13)&	$-$443.4$\pm	7$&	4.9&		&	&	r	\\
B196-G246&	0:43:48.57&	40:42:36.8&	0.23&	$-$312.8$\pm	7$&	5.3&	$-$1.1$\pm	0.1$&	(14)&		\\
B197-G247&	0:43:49.72&	41:30:10.1&	0.30&	$-$60.0$\pm	8$&	5.3&	$-$0.3$\pm	0.1$&	14.0&	e	\\
B199-G248&	0:43:49.83&	40:58:14.8&	0.13&	$-$367.1$\pm	8$&	5.2&	$-$1.5$\pm	0.1$&	(14)&		\\
B198-G249&	0:43:50.11&	41:31:52.6&	0.22&	$-$2.3$\pm	8$&	5.1&	$-$1.0$\pm	0.2$&	(14)&	e	\\
B200&	0:43:50.44&	41:29:22.9&	0.34&	$-$361.3$\pm	34$&	5.1&	$-$1.4$\pm	0.3$&	(14)&	e	\\
B201-G250&	0:43:52.83&	41:09:58.1&	0.12&	$-$700.1$\pm	5$&	5.7&	$-$1.1$\pm	0.1$&	(14)&		\\
B202-G251&	0:43:54.69&	41:00:32.5&	0.20&	$-$342.6$\pm	7$&	5.1&	$-$1.2$\pm	0.1$&	(14)&		\\
B203-G252&	0:43:55.83&	41:32:35.1&	0.17&	$-$229.2$\pm	8$&	5.5&	$-$0.8$\pm	0.1$&	(14)&		\\
B204-G254&	0:43:56.42&	41:22:02.9&	0.13&	$-$356.8$\pm	9$&	5.9&	$-$0.7$\pm	0.1$&	12.1&		\\
B361-G255&	0:43:57.10&	40:14:01.2&	0.12&	$-$341.5$\pm	10$&	5.3&	$-$1.4$\pm	0.2$&	(14)&		\\
B205-G256&	0:43:58.17&	41:24:38.3&	0.12&	$-$369.8$\pm	11$&	6.0&	$-$0.9$\pm	0.1$&	7.7&		\\
B206-G257&	0:43:58.63&	41:30:18.1&	0.10&	$-$190.0$\pm	12$&	6.1&	$-$1.1$\pm	0.1$&	(14)&		\\
B110D-V296&	0:43:59.14&	41:36:41.3&	0.28&	$-$251.5$\pm	12$&	4.9&	$-$1.4$\pm	0.2$&	(14)&		\\
B207-G258&	0:43:59.44&	41:06:10.7&	0.08&	$-$161.2$\pm	12$&	5.2&	$-$1.3$\pm	0.1$&	(14)&		\\
B208-G259&	0:44:00.08&	41:23:11.6&	0.21&	$-$253.5$\pm	10$&	5.0&	$-$0.4$\pm	0.2$&	10.5&		\\
M009&	0:44:00.83&	41:17:12.5&	0.12&	$-$349.2$\pm	20$&	5.0&	$-$1.8$\pm	0.3$&	(14)&	e	\\
B209-G261&	0:44:02.63&	41:25:26.7&	0.13&	$-$461.0$\pm	13$&	5.5&	$-$1.0$\pm	0.1$&	(14)&		\\
B211-G262&	0:44:02.92&	41:20:04.8&	0.14&	$-$140.2$\pm	16$&	5.5&	$-$1.4$\pm	0.1$&	(14)&		\\
B212-G263&	0:44:03.05&	41:04:56.4&	0.16&	$-$401.3$\pm	11$&	6.0&	$-$1.7$\pm	0.1$&	(14)&		\\
B213-G264&	0:44:03.52&	41:30:38.7&	0.17&	$-$566.9$\pm	10$&	5.5&	$-$0.8$\pm	0.1$&	(14)&		\\
B214-G265&	0:44:03.96&	41:26:18.6&	0.15&	$-$274.7$\pm	15$&	5.1&	$-$1.3$\pm	0.1$&	(14)&		\\
B215-G266&	0:44:06.40&	41:31:43.7&	0.14&	$-$150.8$\pm	7$&	5.3&	$-$0.6$\pm	0.1$&	(14)&		\\
B217-G269&	0:44:10.60&	41:23:51.2&	0.13&	$-$15.1$\pm	12$&	5.6&	$-$0.8$\pm	0.1$&	13.9&		\\
M019&	0:44:11.71&	41:23:54.0&	0.12&	$-$260.6$\pm	16$&	4.8&	$-$1.4$\pm	0.2$&	(14)&		\\
B218-G272&	0:44:14.33&	41:19:19.4&	0.17&	$-$220.2$\pm	6$&	6.3&	$-$0.8$\pm	0.1$&	8.7&		\\
B219-G271&	0:44:15.04&	40:56:47.3&	0.17&	$-$503.0$\pm	6$&	5.7&	$-$0.6$\pm	0.1$&	(14)&		\\
B363-G274&	0:44:17.25&	40:33:35.1&	0.13&	$-$374.7$\pm	18$&	5.0&	$-$2.1$\pm	0.2$&	(14)&		\\
B220-G275&	0:44:19.44&	41:30:35.0&	0.13&	$-$275.4$\pm	12$&	5.6&	$-$1.5$\pm	0.2$&	(14)&		\\
B221-G276&	0:44:23.07&	41:33:06.4&	0.26&	$-$457.0$\pm	10$&	5.6&	$-$1.1$\pm	0.1$&	(14)&		\\
B224-G279&	0:44:27.10&	41:28:50.0&	0.15&	$-$139.2$\pm	21$&	6.1&	$-$1.5$\pm	0.1$&	(14)&		\\
B279-D068&	0:44:27.99&	41:44:10.3&	0.29&	$-$114.6$\pm	9$&	4.9&	0.0$\pm	0.2$&	(14)&		\\
B225-G280&	0:44:29.55&	41:21:35.8&	0.12&	$-$158.6$\pm	10$&	6.5&	$-$0.5$\pm	0.1$&	10.7&		\\
B228-G281&	0:44:33.21&	41:41:27.8&	0.24&	$-$437.8$\pm	10$&	5.6&	$-$0.7$\pm	0.1$&	(14)&		\\
B229-G282&	0:44:33.83&	41:38:28.5&	0.11&	$-$39.2$\pm	27$&	5.3&	$-$2.1$\pm	0.2$&	(14)&		\\
B230-G283&	0:44:35.18&	40:57:12.2&	0.12&	$-$575.0$\pm	13$&	5.7&	$-$1.9$\pm	0.1$&	(14)&		\\
B365-G284&	0:44:36.46&	42:17:20.9&	0.19&	$-$60.7$\pm	12$&	5.5&	$-$1.4$\pm	0.1$&	(14)&		\\
B231-G285&	0:44:38.59&	41:27:47.0&	0.15&	$-$305.7$\pm	8$&	5.3&	$-$1.0$\pm	0.1$&	(14)&		\\
B232-G286&	0:44:40.23&	41:15:00.7&	0.21&	$-$192.5$\pm	11$&	6.0&	$-$1.9$\pm	0.1$&	(14)&		\\
B233-G287&	0:44:42.12&	41:43:54.6&	0.17&	$-$72.0$\pm	11$&	5.9&	$-$1.1$\pm	0.1$&	(14)&		\\
B234-G290&	0:44:46.38&	41:29:17.8&	0.23&	$-$202.2$\pm	10$&	5.6&	$-$0.8$\pm	0.1$&	(14)&	e	\\
B366-G291&	0:44:46.72&	42:03:50.5&	0.11&	$-$135.7$\pm	15$&	5.5&	$-$2.0$\pm	0.2$&	(14)&		\\
B283-G296&	0:44:55.37&	41:17:00.2&	0.19&	$-$89.9$\pm	10$&	5.0&	$-$0.8$\pm	0.2$&	(14)&		\\
B235-G297&	0:44:57.93&	41:29:24.0&	0.14&	$-$90.7$\pm	11$&	5.6&	$-$0.9$\pm	0.1$&	10.6&		\\
B236-G298&	0:45:08.90&	40:50:28.6&	0.10&	$-$393.1$\pm	14$&	5.2&	$-$1.5$\pm	0.2$&	(14)&		\\
B237-G299&	0:45:09.22&	41:22:34.6&	0.18&	$-$99.0$\pm	15$&	5.3&	$-$1.8$\pm	0.2$&	(14)&		\\
B370-G300&	0:45:14.39&	41:57:40.8&	0.27&	$-$352.7$\pm	14$&	5.8&	$-$1.6$\pm	0.1$&	(14)&		\\
B238-G301&	0:45:14.67&	41:19:37.1&	0.16&	$-$44.2$\pm	5$&	5.6&	$-$0.7$\pm	0.1$&	(14)&		\\
B239-M74&	0:45:15.6&	41:35:17.2&	(0.12)&	$-$245.9$\pm	11$&	5.3&	$-$0.9$\pm	0.1$&	7.6&		\\
B240-G302&	0:45:25.04&	41:06:22.1&	0.13&	$-$50.4$\pm	7$&	6.1&	$-$1.5$\pm	0.1$&	(14)&		\\
B287&	0:45:28.49&	41:30:04.8&	0.07&	$-$281.4$\pm	34$&	4.9&	$-$1.4$\pm	0.3$&	(14)&		\\
B372-G304&	0:45:33.39&	42:00:24.4&	0.13&	$-$226.9$\pm	8$&	5.5&	$-$1.1$\pm	0.1$&	(14)&		\\
B373-G305&	0:45:41.85&	41:45:33.6&	0.25&	$-$216.2$\pm	7$&	6.0&	$-$0.5$\pm	0.1$&	13.7&		\\
V129-BA4&	0:45:44.69&	41:51:59.4&	0.18&	$-$640.3$\pm	14$&	5.4&	$-$1.4$\pm	0.1$&	(14)&	e	\\
B375-G307&	0:45:45.58&	41:39:42.4&	0.21&	$-$199.0$\pm	7$&	5.3&	$-$0.9$\pm	0.1$&	11.8&		\\
B377-G308&	0:45:48.28&	40:38:04.2&	0.09&	$-$216.6$\pm	10$&	5.3&	$-$1.5$\pm	0.1$&	(14)&		\\
B378-G311&	0:45:57.24&	41:53:31.1&	0.13&	$-$247.5$\pm	15$&	5.1&	$-$1.6$\pm	0.2$&	(14)&		\\
B379-G312&	0:45:58.83&	40:42:31.3&	0.16&	$-$333.2$\pm	6$&	5.8&	$-$0.4$\pm	0.1$&	10.5&		\\
B381-G315&	0:46:06.54&	41:20:58.8&	0.24&	$-$78.2$\pm	12$&	6.0&	$-$1.1$\pm	0.1$&	(14)&		\\
B486-G316&	0:46:08.62&	40:58:03.6&	0.14&	$-$219.7$\pm	15$&	5.2&	$-$1.4$\pm	0.2$&	(14)&		\\
B382-G317&	0:46:10.32&	41:37:40.5&	0.18&	$-$295.4$\pm	11$&	5.3&	$-$1.9$\pm	0.2$&	(14)&		\\
B383-G318&	0:46:11.94&	41:19:41.4&	0.20&	$-$227.6$\pm	6$&	6.1&	$-$0.6$\pm	0.1$&	13.6&		\\
B384-G319&	0:46:21.93&	40:16:59.6&	0.10&	$-$358.4$\pm	7$&	5.8&	$-$0.7$\pm	0.1$&	13.5&		\\
B386-G322&	0:46:27.00&	42:01:52.8&	0.18&	$-$391.8$\pm	11$&	6.0&	$-$1.1$\pm	0.1$&	(14)&		\\
B387-G323&	0:46:33.51&	40:44:13.4&	0.08&	12.1$\pm	14$&	5.3&	$-$2.2$\pm	0.2$&	(14)&		\\
B391-G328&	0:46:58.10&	41:33:56.5&	0.24&	$-$324.2$\pm	10$&	5.4&	$-$1.2$\pm	0.2$&	(14)&		\\
B393-G330&	0:47:01.20&	41:24:06.3&	0.17&	$-$371.6$\pm	14$&	5.4&	$-$0.9$\pm	0.2$&	(14)&		\\
B396-G335&	0:47:25.15&	40:21:42.1&	0.08&	$-$612.4$\pm	28$&	5.2&	$-$2.0$\pm	0.3$&	(14)&		\\
B397-G336&	0:47:27.23&	41:12:10.4&	0.22&	$-$118.8$\pm	14$&	5.7&	$-$1.2$\pm	0.1$&	(14)&		\\
B398-G341&	0:47:57.78&	41:48:45.6&	0.25&	$-$242.7$\pm	13$&	5.3&	$-$0.4$\pm	0.2$&	(14)&		\\
B399-G342&	0:47:59.55&	41:35:28.3&	0.17&	$-$427.8$\pm	27$&	5.3&	$-$1.7$\pm	0.5$&	(14)&		\\
B401-G344&	0:48:08.50&	41:40:41.9&	0.18&	$-$326.1$\pm	21$&	5.5&	$-$1.9$\pm	0.3$&	(14)&		\\
B337D&	0:49:11.20&	41:07:21.0&	0.12&	$-$222.1$\pm	23$&	4.9&	$-$1.7$\pm	0.6$&	(14)&		\\
B403-G348&	0:49:17.62&	41:35:08.1&	0.26&	$-$258.5$\pm	11$&	5.8&	$-$0.9$\pm	0.1$&	(14)&		\\
B405-G351&	0:49:39.80&	41:35:29.7&	0.18&	$-$171.2$\pm	15$&	6.2&	$-$1.2$\pm	0.1$&	(14)&		\\

\enddata
\tablenotetext{a}{Values in parentheses were not derived here. Such values are either the mode of
all clusters, 0.13, or come from \cite{barmby}.}
\tablenotetext{b}{A value of exactly 14 Gyr means the age could not be determined, either because the metallicity
was lower than -0.95 or the data point fell off of the grids shown in Figure \ref{hbhdfem}. 
A blank entry means that we have no blue spectrum or {\it HST} image of the object to estimate the age,
but that the red spectrum or the ground-based image do indicate an old age.  Uncertainties are 2 Gyr for measured ages.
}
\tablenotetext{c}{ ``e'' indicates the spectrum shows weak emission. ``w'' means the spectrum was weak. 
``r'' indicates we have only a red spectrum. For either of the latter two cases, no 
age or metallicity was derived, and it is possible some of these are not in fact old.}
\tablenotetext{d}{The HectoChelle velocity for B070-G133 is significantly different, $-222.9$.}
\tablenotetext{f}{The HectoChelle velocity for B132-NB15 is significantly different, $70.8$.}
\tablenotetext{g}{The HectoChelle velocity for B262 is significantly different, $-340.3$.}
\tablenotetext{h}{The HectoChelle velocity for NB21 is significantly different, $-748.1$.}
\tablenotetext{i}{Velocity for this cluster comes from HectoChelle.}
\tablenotetext{j}{Probably bound to NGC~205.}
\end{deluxetable}

\clearpage
\pagestyle{empty}
\begin{deluxetable}{lllrrrr}
\setlength{\tabcolsep}{0.02in} 
\tabletypesize{\scriptsize}
\tablecolumns{7}
\tablewidth{0pc}
\tablecaption{Young and intermediate age clusters called ``old'' in Paper I \label{revised}} 

\tablehead{\colhead{Object} & \colhead{RA}  &\colhead{Dec} &\colhead{\footnotesize{$E(B-V)$}}  &\colhead{Velocity}   &\colhead{log Mass}  &\colhead{Age\tablenotemark{a}} \\
\colhead{} & \multicolumn{2}{c}{J2000} &\colhead{}    &\colhead{\kms} &\colhead{$M_{\sun}$} &\colhead{Gyr}  
}
\startdata
B305-D024&	0:38:58.85&	40:16:32.1&	(0.13)&	$-$516.9$\pm	 15$&	 4.8&	0.9	\\
B436&	0:39:30.67&	40:18:20.6&	(0.13)&	$-$488.2$\pm	 19$&	 4.7&	1.3	\\
BH09&	0:40:37.15&	40:33:21.9&	&	$-$510.7$\pm	 29$&	&		\\
B335-V013&	0:40:41.67&	40:38:27.9&	(0.13)&	$-$537.9$\pm	 10$&	 4.8&	1.5	\\
B449-V11&	 0:40:42.3&	40:36:04.9&	(0.13)&	$-$551.4$\pm	 15$&	 4.5&	1.3	\\
DAO38&	0:40:47.01&	40:40:57.9&	(0.13)&	$-$537.7$\pm	 22$&	 4.3&	1.2	\\
BH11&	0:40:50.83&	40:40:38.4&	&	$-$549.8$\pm	 25$&	&		\\
KHM31-74&	0:40:52.99&	40:35:19.8&	&	$-$576.5$\pm	 25$&	&		\\
KHM31-77&	0:40:53.69&	40:36:50.8&	&	$-$556.0$\pm	 26$&	&		\\
B247&	0:41:02.27&	41:00:32.0&	(0.13)&	$-$521.1$\pm	  9$&	 4.7&	1.5	\\
LGS04105.6\_410743\tablenotemark{b}&	0:41:05.59&	41:07:42.7&	0.58&	$-$265.9$\pm	 32$&	 4.9&	0.8	\\
B017D&	0:41:10.01&	40:58:10.6&	&	$-$500.0$\pm	 23$&	&		\\
SK036A&	0:41:47.40&	40:51:08.6&	&	$-$571.4$\pm	 40$&	&		\\
B259&	 0:42:19.0&	41:42:13.9&	&	$-$294.2$\pm	 24$&	&		\\
B051D&	0:42:20.56&	41:04:37.7&	&	$-$213.7$\pm	 22$&	&		\\
NB35-AU4&	0:42:34.55&	41:18:40.4&	&	$-$302.3$\pm	 35$&	&		\\
SK068A&	0:43:28.15&	41:00:22.0&	&	$-$285.1$\pm	 33$&	&		\\
M012&	0:44:02.83&	41:21:40.3&	0.28&	$-$220.7$\pm	 14$&	 4.5&	1.1	\\
B277-M22&	0:44:16.90&	41:14:16.0&	(0.13)&	$-$358.1$\pm	  9$&	 4.1&	1.0	\\
M026&	0:44:20.24&	41:27:20.0&	&	&		&		\\
B112D-M27&	0:44:21.23&	41:19:09.8&	0.22&	$-$290.6$\pm	 16$&	 4.5&	1.0	\\
M040&	0:44:31.51&	41:27:55.2&	&	$-$204.5$\pm	 30$&	&		\\
M043&	0:44:34.36&	41:23:11.5&	&	$-$199.4$\pm	 39$&	&		\\
M045&	 0:44:36.4&	41:35:32.9&	&	$-$178.1$\pm	 29$&	&		\\
M047&	 0:44:37.8&	41:28:51.9&	(0.13)&	 $-$36.2$\pm	 20$&	 4.3&	1.3	\\
B281-G288&	0:44:42.85&	41:20:08.6&	(0.08)&	$-$215.3$\pm	  7$&	 4.8&	1.1	\\
B368-G293&	0:44:47.80&	41:51:09.2&	&	&		&		\\
M053&	0:44:57.29&	41:48:02.0&	(0.13)&	 $-$67.6$\pm	 19$&	 4.2&	1.0	\\
B257D-D073\tablenotemark{c}&	0:44:59.49&	41:54:46.9&	0.37&	$-$124.2$\pm	 17$&	 4.2&	0.2	\\
M057&	0:45:02.75&	41:47:02.4&	&	 $-$80.4$\pm	 39$&	&		\\
M058&	0:45:03.36&	41:40:05.5&	&	$-$168.7$\pm	 26$&	&		\\
KHM31-264&	0:45:05.79&	41:35:42.5&	&	 $-$85.9$\pm	 40$&	&		\\
B476-D074&	0:45:07.18&	41:40:31.1&	(0.13)&	$-$151.1$\pm	 20$&	 4.5&	1.2	\\
M070&	0:45:11.79&	41:40:19.8&	(0.13)&	$-$136.4$\pm	 19$&	 4.0&	1.2	\\
M081&	0:45:22.29&	41:47:57.0&	0.27&	 $-$93.4$\pm	 21$&	 3.7&	0.9	\\
M105&	0:45:49.70&	41:39:26.0&	(0.13)&	$-$156.8$\pm	 14$&	 3.9&	1.2	\\
DAO84\tablenotemark{d}&	0:45:52.33&	41:42:49.2&	0.27&	$-$170.6$\pm	 27$&	 3.7&	0.2	\\

\enddata
\tablenotetext{a}{Clusters without ages have spectra too poor to determine one, but the age
is expected to be younger than 3 Gyr based on the continuum shape or the cluster's {\it HST} image. Uncertainties in age are roughly a factor of 2.}
\tablenotetext{b}{New cluster not listed in Paper I. }
\tablenotetext{c}{The coordinates for B257D-D073 have been corrected from Paper I. }
\tablenotetext{d}{The velocity for DAO84 in Paper I was incorrect due to an automatic misidentification
of emission lines, and thus this young cluster was
erroneously identified as a background galaxy. We thank Paul Hodge for helping
us identify this error. No other object listed in the tables of Paper I has this particular error.}
\end{deluxetable}

\clearpage
\pagestyle{empty}
\begin{deluxetable}{llll}
\tablecolumns{4}
\tablewidth{0pc}
\tablecaption{Comparison of [Fe/H] derived from spectra and {\it HST} CMDs \label{compare_feh}} 
\tablehead{\colhead{Object} & \colhead{Spectra}  &\colhead{CMD} &\colhead{Ref\tablenotemark{a}} }
\startdata
B006-G058&	$-$0.5$\pm	0.1$&	$-$0.6&	b	\\
B008-G060&	$-$0.8$\pm	0.1$&	$-$1.4, $-$1.0&	here, e	\\
B010-G062&	$-$1.8$\pm	0.2$&	$-$2.1, $-$1.8&	here, e	\\
B012-G064&	$-$1.7$\pm	0.1$&	$-$1.8&	b	\\
B023-G078&	$-$0.7$\pm	0.1$&	$-$0.7, $-$0.7&	a, e	\\
B027-G087&	$-$1.3$\pm	0.1$&	$-$1.4&	b	\\
B045-G108&	$-$0.9$\pm	0.1$&	$-$0.6&	b	\\
B057-G118&	$-$2.1$\pm	0.3$&	$-$2.3&	here	\\
B058-G119&	$-$1.1$\pm	0.1$&	$-$1.3&	b	\\
B088-G150&	$-$1.8$\pm	0.1$&	$-$1.9&	e	\\
B158-G213&	$-$0.8$\pm	0.1$&	$-$0.7, $-$0.9&	a, e	\\
B220-G275&	$-$1.5$\pm	0.2$&	$-$1.8&	e	\\
B224-G279&	$-$1.5$\pm	0.1$&	$-$1.8&	e	\\
B225-G280&	$-$0.5$\pm	0.1$&	$-$0.6, $-$0.4, $-$0.2, $-$0.6&	a,b,c,e	\\
B233-G287&	$-$1.1$\pm	0.1$&	$-$1.6&	b	\\
B240-G302&	$-$1.5$\pm	0.1$&	$-$1.4, $-$1.9&	b,d	\\
B311-G033&	$-$1.9$\pm	0.2$&	$-$1.6&	b	\\
B338-G076&	$-$1.1$\pm	0.1$&	$-$1.3&	b	\\
B343-G105&	$-$1.6$\pm	0.2$&	$-$1.5&	b	\\
B366-G291&	$-$2.0$\pm	0.2$&	$-$1.9, $-$1.8&	here, e	\\
B379-G312&	$-$0.4$\pm	0.1$&	$-$0.5, $-$0.6&	b,d	\\
B384-G319&	$-$0.7$\pm	0.1$&	$-$0.7&	b	\\
B386-G322&	$-$1.1$\pm	0.1$&	$-$1.2&	b	\\
B405-G351&	$-$1.2$\pm	0.1$&	$-$1.7&	b	\\
\enddata
\tablenotetext{a}{References - a: \cite{fuentes}, b: \cite{rich} Table 5, Column 6, c: \cite{stephens}, d: \cite{holland}, e: \cite{perina}}
\end{deluxetable}

\end{document}